\documentclass[10pt,aps,prb,twocolumn,superscriptaddress,showpacs,longbibliography,floatfix]{revtex4-2}
\usepackage{amssymb}
\usepackage{graphicx}
\usepackage{amsmath}
\usepackage{mathtools}
\usepackage[export]{adjustbox}
\usepackage{epsfig}
\usepackage{times}
\usepackage{color}
\usepackage{subfigure}
\usepackage{setspace}
\usepackage{natbib}
\usepackage[normalem]{ulem}
\usepackage{cancel}
\usepackage{soul}
\usepackage{physics}
\usepackage{multirow}

\newcounter{algorithm}
\renewcommand{\thealgorithm}{\Roman{algorithm}}
\newcommand{\algotitle}[1]{%
  \refstepcounter{algorithm}%
  \noindent ALGORITHM~\thealgorithm.\ #1\par\smallskip
}
\usepackage{algpseudocode}

\usepackage[pdftex]{hyperref}

\usepackage{xcolor}
\usepackage{xparse,xcoffins}
\ExplSyntaxOn
\NewCoffin\imagecoffin
\NewCoffin\labelcoffin
\keys_define:nn { miguel/label }
 {
  label   .tl_set:N = \l_miguel_label_tl,

  labelbox .bool_set:N = \l_miguel_label_box_bool,
  labelbox .default:n = true,
  fontsize .tl_set:N = \l_miguel_label_size_tl,
  fontsize .initial:n = \footnotesize,
  pos .choice:,
  pos/nw .code:n = \tl_set:Nn \l_miguel_label_pos_tl { left,up },
  pos/ne .code:n = \tl_set:Nn \l_miguel_label_pos_tl { right,up },
  pos/sw .code:n = \tl_set:Nn \l_miguel_label_pos_tl { left,down },
  pos/se .code:n = \tl_set:Nn \l_miguel_label_pos_tl { right,down },
  pos/n .code:n = \tl_set:Nn \l_miguel_label_pos_tl { hc,up },
  pos/w .code:n = \tl_set:Nn \l_miguel_label_pos_tl { left,vc },
  pos/s .code:n = \tl_set:Nn \l_miguel_label_pos_tl { hc,down },
  pos/e .code:n = \tl_set:Nn \l_miguel_label_pos_tl { right,vc },
  pos .initial:n = nw,
  unknown .code:n   = \clist_put_right:Nx \l_miguel_label_clist
                       { \l_keys_key_tl = \exp_not:n { #1 } }
 }
\clist_new:N \l_miguel_label_clist
\box_new:N \l_miguel_label_box
\box_new:N \l_miguel_label_image_box
\NewDocumentCommand{\xincludegraphics}{O{}m}
 {
  \group_begin:
  \tl_clear:N \l_miguel_label_tl
  \clist_clear:N \l_miguel_label_clist
  \keys_set:nn { miguel/label } { #1 }
  \tl_if_empty:NTF \l_miguel_label_tl
   {
    \miguel_includegraphics:Vn \l_miguel_label_clist { #2 }
   }
   {
    \SetHorizontalCoffin\imagecoffin
     {
      \miguel_includegraphics:Vn \l_miguel_label_clist { #2 }
     }
    \SetHorizontalCoffin\labelcoffin
     {
      \raisebox{\depth}
       {
        \bool_if:NTF \l_miguel_label_box_bool
         { \fcolorbox{white}{white}{\l_miguel_label_size_tl\l_miguel_label_tl} }
         { \l_miguel_label_size_tl\l_miguel_label_tl }
       }
     }
    \SetVerticalPole\imagecoffin{left}{3pt+\CoffinWidth\labelcoffin/2}
    \SetVerticalPole\imagecoffin{right}{\Width-3pt-\CoffinWidth\labelcoffin/2}
    \SetHorizontalPole\imagecoffin{up}{\Height-3pt-\CoffinHeight\labelcoffin/2}
    \SetHorizontalPole\imagecoffin{down}{3pt+\CoffinHeight\labelcoffin/2}
    \use:x{\JoinCoffins\imagecoffin[\l_miguel_label_pos_tl]\labelcoffin[vc,hc]} 
    \TypesetCoffin\imagecoffin
   }
   \group_end:
 }
\NewDocumentCommand{\setlabel}{m}
 {
  \keys_set:nn { miguel/label } { #1 }
 }
\cs_new_protected:Nn \miguel_includegraphics:nn
 {
  \includegraphics[#1]{#2}
 }
\cs_generate_variant:Nn \miguel_includegraphics:nn { V }
\ExplSyntaxOff

\usepackage{xcolor}
\definecolor{burntorange}{HTML}{C45A00}

\begin{document}

\title{Learning Spin Hamiltonians from Interacting Magnon Spectra}

\author{Ang-Kun Wu}
\email[Corresponding author: ]{angkunwu@gmail.com}
\affiliation{Department of Physics and Astronomy, University of Tennessee, Knoxville, Knoxville,
Tennessee, 37996, USA}
\author{David A. Dahlbom}
\affiliation{Neutron Scattering Division, Oak Ridge National Laboratory, Oak Ridge, Tennessee 37831,
USA}
\author{Kipton Barros}
\affiliation{Theoretical Division and CNLS, Los Alamos National Laboratory, Los Alamos,
New Mexico 87545, USA}
\author{Cristian D. Batista}
\affiliation{Department of Physics and Astronomy, University of Tennessee, Knoxville, Knoxville,
Tennessee, 37996, USA}
\affiliation{Neutron Scattering Division and Shull-Wollan Center, Oak Ridge National Laboratory,
Oak Ridge, Tennessee 37831, USA}
\author{Hao Zhang}
\email[Corresponding author: ]{hzhang68@utk.edu}
\affiliation{Department of Physics and Astronomy, University of Tennessee, Knoxville, Knoxville,
Tennessee, 37996, USA}

\begin{abstract}
A neural network that learns an inverse map inherits the approximations of the forward solver used to generate its training data. We demonstrate this in the inference of spin Hamiltonians from synthetic neutron-scattering spectra and show how a fast many-body solver can overcome the limitation of linear spin-wave theory (LSWT). We consider a family of anisotropic triangular-lattice spin Hamiltonians at fields just above saturation, where magnon interactions shift and broaden the spectrum in ways that LSWT cannot capture. Using the truncated Hilbert-space ladder (THL) method, we compute the dynamical spin structure factor (DSSF) nonperturbatively within the one- and two-magnon space. THL is fast enough to generate large training sets of DSSF spectra, enabling us to train neural networks that accurately predict Hamiltonian parameters.  By empirically benchmarking several neural-network architectures, we find that a one-dimensional convolutional network performs best.
A learned forward decoder then reconstructs the DSSF from the predicted parameters, providing a rapid self-consistency check and a visualization of the predicted response. When interactions are strong, networks trained on LSWT and THL spectra infer markedly different parameters. Analysis of the Fisher information matrix shows that the measurement protocol, including which spin-correlation components are resolved and the momentum cut, determines which couplings can be identified. Combining machine learning with fast many-body theory is therefore essential for quantitative Hamiltonian inference from scattering data.
\end{abstract}

\maketitle

\section{Introduction\label{sec:intro}}

\begin{figure*}[t]
\begin{center}
\includegraphics[width=0.95\linewidth]{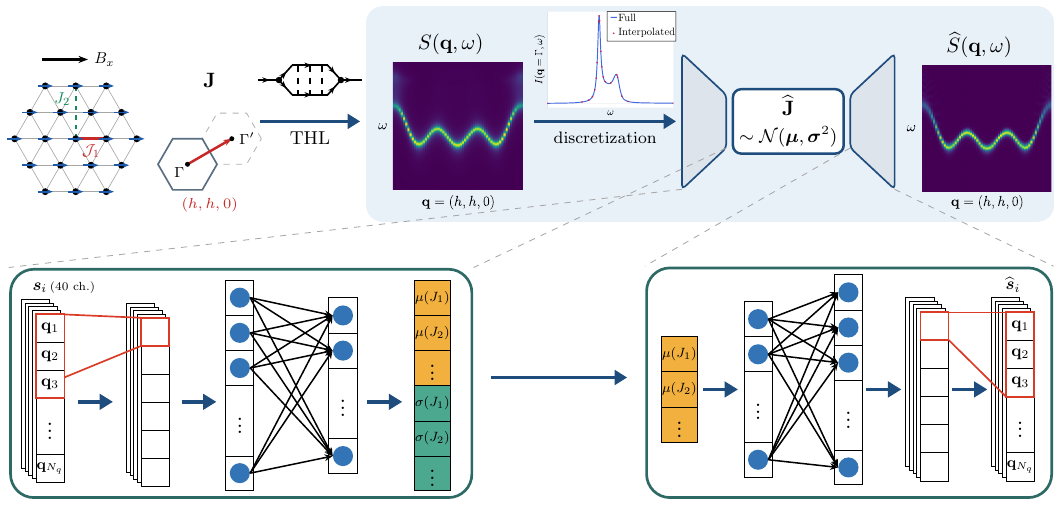}
\caption{\label{fig:schematic}
From the DSSF to the Hamiltonian and back. Spin-$1/2$ moments on a triangular lattice interact through an anisotropic nearest-neighbor exchange $\mathcal{J}_1$, a $3\times3$ matrix (red solid bond), and an isotropic next-nearest-neighbor exchange $J_2$ (green dashed bond). A magnetic field $B_x$ polarizes them along $x$ (blue arrows). For given parameters $\mathbf{J}=(J_1,J_{zz},J_{\pm\pm},J_{z\pm},J_2)$, we compute the DSSF $S(\mathbf{q},\omega)$ along $\mathbf{q}=(h,h,0)$, from $\Gamma$ to $\Gamma'$, with the truncated Hilbert-space ladder (THL) method, whose ladder diagram describes the repeated scattering of magnon pairs, or with LSWT. The shaded region contains the inversion and the reconstruction. At each momentum, adaptive discretization keeps a set of knots (red dots) from which the spectrum (blue line) is approximated by linear interpolation. The inverse network maps these knots to a Gaussian distribution of Hamiltonian parameters
with uncertainties, and a learned forward decoder maps the mean $\boldsymbol{\mu}$ back to a reconstructed DSSF $\widehat{S}(\mathbf{q},\omega)$, which tests the consistency of the inversion. The bottom row shows the two networks. The input $\boldsymbol{s}_i$ of the inverse network is defined along the momentum path $\mathbf{q}_1,\ldots,\mathbf{q}_{N_k}$, with 40 channels at each momentum holding the energies and intensities of 20 knots, as an example. %
The inverse network applies a one-dimensional convolution along the momentum axis and uses a fully connected head to predict the mean and standard deviation of each coupling. The forward decoder takes the inferred mean couplings as input. Interpolating its output knots reconstructs the DSSF.
}
\end{center}
\end{figure*}

How much can we learn about the microscopic interactions in a quantum material by measuring its excitations? Even a precisely measured spectrum can leave several Hamiltonians consistent with the data. Which couplings we can determine depends both on the observables we measure and on which features of the measured response our theory can predict. Two spin Hamiltonians, for example, may yield nearly identical one-magnon dispersions and peak intensities yet predict different incoherent two-magnon continua. The energy at which the continuum begins and the way its spectral weight varies with momentum and energy can then distinguish the two models. To exploit this information, the physics simulation (forward solver) must go beyond the single-particle description of linear spin-wave theory. Here we introduce a solver that treats magnon--magnon interactions and is fast enough to generate the spectra needed to train a neural network for Hamiltonian inference. The same calculations let us ask, before any data are collected, which momentum cuts and polarization channels will best distinguish the candidate Hamiltonians (Fig.~\ref{fig:schematic}).

Inelastic neutron scattering (INS) probes the dynamical spin structure factor (DSSF), which describes how spin correlations are distributed in momentum and energy~\cite{boothroyd2020principles,squires1996introduction}. For bilinear spin Hamiltonians, the full set of equal-time spin correlations can determine a nondegenerate ground-state wave function~\cite{Quintanilla2022} and, at a known finite temperature, the exchange couplings themselves~\cite{Murta2022}. An INS experiment samples a finite range of momenta and energies, and the scattering geometry and polarization analysis determine which components of the correlation tensor it resolves~\cite{Moon1969Polarization,Violini2014}. Conventionally, one then fits a Hamiltonian to these data~\cite{scheie2020multiphase}, often using linear spin-wave theory (LSWT) to calculate the response~\cite{Toth2015SpinW,Dahlbom2025Sunny}. If changes in two couplings compensate each other over the measured range, the fit can develop a broad valley of nearly equivalent Hamiltonians~\cite{Transtrum2010NonlinearFits}. The variation of the predicted spectra along this valley can then reveal which additional momentum cuts or polarization channels would distinguish the competing Hamiltonians.

Simulation-based inference offers a way to compare proposed measurements before the experiment~\cite{Cranmer2020SBI}. For each measurement protocol, a forward solver generates spectra for Hamiltonians with known couplings. One can then train a network to infer those couplings and test it on Hamiltonians withheld from training. Comparing the errors for each coupling reveals what one gains by changing the momentum cut or resolving an additional polarization channel.

Neural networks have been used throughout quantum many-body physics~\cite{Carleo2019MLPhysics} to identify phases~\cite{carrasquilla2017machine}, represent wave functions~\cite{Carleo2017,wu2026compact}, model quantum geometry~\cite{wu2026modeling}, and reconstruct states from measurements~\cite{torlai2018neural}. In magnetism, they have helped constrain exchange interactions in spin ice and $\alpha$-RuCl$_3$~\cite{Samarakoon2020,Samarakoon2022Pressure,Samarakoon2022RuCl3}, distinguish competing exchange models~\cite{Butler2021}, connect simulated and measured spectra~\cite{Anker2023}, and refine Hamiltonian parameters against INS data~\cite{Chitturi2023}. Related approaches infer magnetic interactions from powder diffraction~\cite{desai2026machine} and terahertz coherent spectra~\cite{Mootz2026THz}. Luo \textit{et al.} trained a network to infer eight exchange parameters from the spectrum of a field-polarized honeycomb Heisenberg model~\cite{luo2026approaching}. In that setting, LSWT gives the exact one-magnon spectrum. When a magnon hybridizes with a two-magnon continuum, the exchange couplings can affect the peak energy and lifetime as well as the distribution of spectral weight throughout the continuum. Learning from these features requires a solver that can resolve them across an entire training set.

This issue arises just above saturation in anisotropic triangular-lattice magnets such as the Yb-based delafossites~\cite{Xie2023CsYbSe2,Dai2021NaYbSe2,Scheie2024NaYbSe2,Scheie2024KYbSe2}, whose neutron spectra have been analyzed using nonlinear spin-wave theory~\cite{Scheie2024Nonlinear}. Anisotropic exchange can mix a single magnon with a pair, shifting the magnon energy, transferring spectral weight to the two-magnon continuum, and allowing decay where the two overlap. Many-body calculations already inform inference in \textit{ab initio} nuclear physics~\cite{Hu2022Pb208}, x-ray spectroscopy of local electronic excitations~\cite{Lueder2021,Lajer2025RIXS,Klein2026RIXS}, and tunneling spectroscopy of a molecular magnet~\cite{Koch2025Triplon}. For an extended magnet, a calculation feasible for one set of exchange parameters can become prohibitive when repeated across the training set. In the dilute regime considered here, the dominant processes can be treated within the one- and two-magnon sectors~\cite{Galitskii1958,Beliaev1958b}. Truncated Hilbert-space exact diagonalization (THED) treats these sectors nonperturbatively on finite clusters~\cite{Bai2023HeavyMagnons,Legros2021FeI2,Zhang2026THED}. The truncated Hilbert-space ladder (THL) method, introduced here, sums repeated scattering of the magnon pair and retains its coupling to the single magnon in the thermodynamic limit~\cite{SunnyLadderImplementation}. For the calculations reported here, THL is more than an order of magnitude faster than THED, making it practical to sample nearly $320{,}000$ Hamiltonians for the five-parameter model. To our knowledge, this is the first use of a nonperturbative solver for magnon--magnon interactions to train a neural network to infer a spin Hamiltonian from simulated INS spectra.

How much can the inferred Hamiltonian change when the forward solver omits magnon interactions? We test this by generating a synthetic neutron spectrum with THL for a Hamiltonian whose exchange couplings are known. The input Hamiltonian has $J_{zz}/J_1=0.25$. From the energies of the main peaks, a network trained on LSWT dispersions infers $J_{zz}/J_1=-0.32$, reversing the sign of the exchange along $z$ (Fig.~\ref{fig:main_result}). Despite this reversal, the LSWT dispersion calculated with the inferred couplings follows the peak energies with a root-mean-square deviation of only $0.05|J_1|$. A direct least-squares fit using LSWT gives nearly the same answer. In both cases, LSWT mimics the peak shifts caused by magnon interactions by changing the exchange couplings. Training on THL spectra addresses this source of error because the forward solver predicts both the shifted peaks and the incoherent two-magnon continuum for each set of couplings. On clean synthetic spectra withheld from training, the THL-trained convolutional network recovers the four exchange parameters with a mean absolute relative error of about $2\%$.

Even with an accurate forward solver, two Hamiltonians can give the same neutron spectrum in a given scattering geometry. We use the Fisher information matrix to measure how sensitively a proposed measurement responds to small changes in the couplings. Direct comparison of simulated spectra also reveals ambiguities between distinct Hamiltonians. In our five-parameter model, reversing the sign of the off-diagonal exchange $J_{z\pm}$ leaves the unpolarized spectrum [Eq.~\eqref{eq:unpolarized_dssf}] along $(h,h,0)$ unchanged. The sign is, however, visible in the cross-correlations $S_{yz}+S_{zy}$ and $S_{xz}+S_{zx}$, both of which reverse sign with $J_{z\pm}$. The first is a one-magnon correlation already present in LSWT. The second arises only from the mixing of one- and two-magnon states. These correlations acquire measurable intensity in an unpolarized neutron scattering experiment
outside of the $(h,h,0)$ plane. In an idealized experimental geometry, changing the cut from $(h,h,0)$ to $(h,h,1)$ reduces the mean squared error in the five inferred couplings from $7.7\times10^{-2}$ to $1.4\times10^{-5}$ on held-out synthetic data. With a field along a nearest-neighbor bond, as considered here, a symmetry of the model confines the sign information to cuts in which the momentum transfer has components along both the field and perpendicular to the plane, such as $(2h,-h,1)$.  This cut, which is experimentally accessible with a horizontal-field magnet, can also resolve the sign, and we give an estimate of how much measurement time is required to resolve the correlation relative to the idealized cut along $(h,h,1)$ (Sec. \ref{sec:identifiable}). Through this example we demonstrate how solving the inverse problem on candidate scans before beam time enables one to choose a measurement geometry and exposure time that resolves the exchange coupling of interest.

\section{Model and neutron-scattering observables\label{sec:model}}

\subsection{Model Hamiltonian}

We consider a spin-$1/2$ system on a triangular lattice. The model under consideration is motivated by a family of spin-$1/2$ $J_1$--$J_2$ triangular-lattice compounds, $A\mathrm{YbSe}_2$ (with $A=\mathrm{Na},\mathrm{K},\mathrm{Cs}$), which reside near a quantum critical regime separating magnetically ordered states from a putative spin-liquid ground state at zero field~\cite{Xie2023CsYbSe2,Dai2021NaYbSe2,Scheie2024NaYbSe2,Scheie2024KYbSe2}. We take $\mathbf{a}_1=a(1, 0, 0)$, $\mathbf{a}_2=a(-1/2,\sqrt{3}/2, 0)$ and $\mathbf{c}=c(0,0,1)$ as lattice vectors; momenta will always be given in reciprocal lattice units of this cell. The Hamiltonian is
\begin{equation}
    \label{eq:model_hamiltonian}
    H =
    \sum_{\langle i,j\rangle}
    \mathbf{S}_i^{\mathsf T}\mathcal{J}_{ij}\mathbf{S}_j
    + J_2\sum_{\langle\langle i,j\rangle\rangle}
    \mathbf{S}_i\cdot\mathbf{S}_j
    - B_x\sum_j S_j^x ,
\end{equation}
where $\langle i,j\rangle$ and $\langle\langle i,j\rangle\rangle$ denote nearest-neighbor and next-nearest-neighbor bonds within each triangular layer.   $B_x$ denotes the Zeeman energy associated with a magnetic field applied along the $x$ direction and therefore has the same units as the exchange couplings. The spin operator is $\mathbf{S}_i=(S_i^x,S_i^y,S_i^z)^{\mathsf T}$, and the nearest-neighbor exchange tensor $\mathcal{J}_{ij}$ on a chosen reference bond takes the form
\begin{equation}
    \label{eq:exchange_matrix}
    \mathcal{J}_1 =
    \begin{pmatrix}
    J_1+2J_{\pm\pm} & 0 & J_{z\pm}\\
    0 & J_1-2J_{\pm\pm} & 0\\
    J_{z\pm} & 0 & J_{zz}
    \end{pmatrix}.
\end{equation}
Here, $J_1$ sets the overall nearest-neighbor exchange scale, $J_{zz}$ couples the $z$ components of neighboring spins, and the ratio $\Delta=J_{zz}/J_1$ measures the axial exchange anisotropy. The lattice symmetry permits two bond-directional exchange anisotropies, parametrized by $J_{\pm\pm}$ and $J_{z\pm}$. The isotropic next-nearest-neighbor exchange coupling is denoted by $J_2$. We collect these five couplings, each with units of energy, in the parameter vector $\mathbf{J}=(J_1,J_{zz},J_{\pm\pm},J_{z\pm},J_2)$.

Equation~\eqref{eq:exchange_matrix} gives the exchange tensor $\mathcal{J}_1$ on a chosen reference bond. Rotations by $\pm 2\pi/3$ about the crystal's threefold axis yield the tensors $\mathcal{J}_{ij}$ for the other two nearest-neighbor bond orientations~\cite{Li2016Anisotropic}. 
The reference bond of Eq.~\eqref{eq:exchange_matrix} lies along $\mathbf a_1\parallel\hat x$, and the field is applied along $\hat x$. The momentum direction $(h,h,0)$ is then parallel to the nearest-neighbor bond $\mathbf a_1+\mathbf a_2$, at $60^\circ$ to the field, so the in-plane spectra correspond to a horizontal-field geometry with $\mathbf c$ vertical.

We study the excitation spectrum in the field-polarized phase of Eq.~\eqref{eq:model_hamiltonian}. At high fields, quantum fluctuations are weak, and linear spin-wave theory (LSWT) accurately describes the spectrum. Just above the saturation field, however, exchange interactions that break $U(1)$ spin-rotation symmetry about the field direction can strongly modify the spectrum. These interactions violate magnon-number conservation and couple single-magnon excitations to the two-magnon continuum, renormalizing magnon energies and allowing decay where the single-magnon branch overlaps the continuum. Such effects lie beyond LSWT, which treats magnons as noninteracting quasiparticles.

In the $A\mathrm{YbSe}_2$ family, fields just above saturation are accessible in inelastic neutron scattering experiments~\cite{Xie2023CsYbSe2,Dai2021NaYbSe2,Scheie2024NaYbSe2,Scheie2024KYbSe2}. We therefore focus on $B_x=B_\mathrm{sat}+|J_1|$, where $B_\mathrm{sat}$ denotes the classical saturation field. The offset $|J_1|$ keeps the single-magnon gap open even after its downward renormalization by quantum fluctuations. The ground state then contains only a small density of virtual magnons, which motivates truncating the excitation space to the one- and two-magnon sectors~\cite{Galitskii1958,Beliaev1958b}. We treat interactions within this space nonperturbatively using the truncated-Hilbert-space ladder (THL) method described in Sec.~\ref{sec:thl}.

\subsection{Observables and the inverse problem}

The dynamical spin structure factor (DSSF) is the Fourier transform of the space- and time-dependent spin correlation function. For a system with $N_s$ sites, its tensor components are
\begin{equation}\label{eq:dssf}
  \begin{split}
    S_{\alpha\beta}(\mathbf{q},\omega)
    &= \frac{1}{2\pi N_s}
    \sum_{i,j}
    e^{-i\mathbf{q}\cdot(\mathbf{r}_i-\mathbf{r}_j)}
    \\
    &\quad\times
    \int_{-\infty}^{\infty}dt\,e^{i\omega t}
    \left\langle S_i^\alpha(t)S_j^\beta(0)\right\rangle,
  \end{split}
\end{equation}
where $\alpha,\beta\in\{x,y,z\}$ and $\langle\cdots\rangle$ denotes a ground-state expectation value.

Inelastic neutron scattering (INS) probes fluctuations of the magnetic moment perpendicular to the momentum transfer $\mathbf{q}$. For a scalar $g$ factor, the unpolarized magnetic intensity is proportional to
\begin{equation}\label{eq:unpolarized_dssf}
  S_{\perp}(\mathbf{q},\omega)
  =
  \sum_{\alpha,\beta}
  \left(
    \delta_{\alpha\beta}
    -\widehat q_\alpha\widehat q_\beta
  \right)
  S_{\alpha\beta}(\mathbf{q},\omega),
\end{equation}
where $\widehat{\mathbf{q}}=\mathbf{q}/|\mathbf{q}|$~\cite{boothroyd2020principles}. The proportionality factor includes the kinematic prefactor and the magnetic form factor. Depending on the scattering geometry, polarization analysis can separate contributions that are summed in the unpolarized intensity~\cite{zaliznyak2014neutron}. The inverse problem is to infer the exchange parameters $\mathbf{J}$ from the measured spectra and to determine how tightly the available polarization channels constrain each coupling.

We solve this inverse problem with neural networks trained on synthetic scattering data (top row of Fig.~\ref{fig:schematic}). The measurement protocol specifies the sampled momenta and energies, the scattering geometry, and the polarization channels. For a fixed protocol, the forward map assigns a predicted scattering response to each set of exchange parameters,
\begin{equation}
  \label{eq:forward_map}
  \mathcal{F}:\mathbf{J}
  \longmapsto S(\mathbf{q},\omega;\mathbf{J}),
\end{equation}
where $S$ denotes the selected unpolarized response or set of polarized components. We consider two models. In the four-parameter model, we fix $J_{z\pm}=0$ and vary $J_1$, $J_{zz}$, $J_{\pm\pm}$, and $J_2$. In the five-parameter model, we also vary $J_{z\pm}$. We compute the synthetic spectra using LSWT and THL, as described in Sec.~\ref{sec:forward}. The forward calculations omit magnetic form factors, detailed instrumental resolution, and correlated background contributions. %

In the forward calculations, we express all energies, including the Zeeman energy $B_x$, in units of $|J_1|$. We set $J_1=\pm1$ when generating the training data, with $J_1=-1$ for ferromagnetic and $J_1=+1$ for antiferromagnetic nearest-neighbor exchange. When inferring the Hamiltonian parameters from the DSSF, however, we treat $J_1$ as a free parameter that sets the overall energy scale. When comparing the inferred parameters between different solvers, we rescale them by the inferred $J_1$ to account for this freedom. Because $B_x=B_\mathrm{sat}(\mathbf J)+|J_1|$ is set separately for each Hamiltonian, the field is not an independent input: the network is effectively given the distance to saturation, which is itself a function of the couplings. In an experiment the field is fixed in absolute units, so a network intended for a specific measurement should be trained at the experimental field, or conditioned on it.

\section{Forward solvers}\label{sec:forward}
\label{sec:solver}

We use LSWT and THL to generate synthetic DSSF datasets for the triangular-lattice Hamiltonian of Sec.~\ref{sec:model}. Both calculations use the field $B_x=B_\mathrm{sat}+|J_1|$ in the polarized phase. LSWT provides an inexpensive description of noninteracting magnons. THL couples the one- and two-magnon sectors and sums repeated two-magnon scattering to all orders in the thermodynamic limit. This nonperturbative treatment remains efficient enough to generate large training datasets. By training and evaluating inverse models on spectra from both solvers, we assess how interaction-induced shifts in excitation energies and redistribution of spectral weight affect the inferred exchange parameters.

\subsection{Linear spin-wave theory}

We use the Holstein--Primakoff transformation~\cite{HolsteinPrimakoff1940} to expand the Hamiltonian about the field-polarized state to quadratic order in the boson operators. A bosonic Bogoliubov transformation then yields the magnon energies and the spin matrix elements that enter the DSSF~\cite{Colpa1978}. The polarized triangular lattice has one spin per magnetic unit cell and therefore a single magnon band $E(\mathbf{q})$. For the LSWT dataset, we retain the one-magnon contribution to the DSSF, which before numerical broadening takes the form
\begin{equation}\label{eq:lswt_dssf}
  S_{\alpha\beta}^{\mathrm{LSWT}}(\mathbf{q},\omega)
  = A_{\alpha\beta}(\mathbf{q})\,
    \delta\!\left[\omega-E(\mathbf{q})\right],
\end{equation}
where $A_{\alpha\beta}(\mathbf{q})$ is the corresponding spectral weight. We use \texttt{Sunny.jl} to compute $E(\mathbf{q})$ and $A_{\alpha\beta}(\mathbf{q})$ at each sampled momentum~\cite{Dahlbom2025Sunny,SunnyRepository}.

\subsection{Nonperturbative methods in a truncated Hilbert space}
\label{sec:thl}

When a single magnon hybridizes with the two-magnon continuum, its energy and lifetime depend on how the two magnons scatter from one another. The truncated Hilbert-space ladder (THL) method retains this hybridization and sums repeated scattering within the pair to all orders through a two-magnon $T$ matrix~\cite{SalpeterBethe1951,GalitskiiMigdal1958,Galitskii1958,Beliaev1958a,Beliaev1958b}. The same ladder resummation has long been used to describe two-magnon bound states in magnets~\cite{Wortis1963,Silberglitt1968}. Within the one- and two-magnon subspace, THL solves the same projected Hamiltonian as truncated Hilbert-space exact diagonalization (THED)~\cite{Legros2021FeI2,Bai2023HeavyMagnons,Zhang2026THED}. THED obtains the excitation spectrum by diagonalizing this Hamiltonian on finite clusters, whereas THL evaluates the interacting response directly in the thermodynamic limit.

At $B_x=B_\mathrm{sat}+|J_1|$, the single-magnon excitation remains gapped and the polarized ground state contains a low density of virtual magnons (Sec.~\ref{sec:model}). This dilute regime motivates a description in which an excitation of total momentum $\mathbf{q}$ is a superposition of a single Bogoliubov magnon at $\mathbf{q}$ and magnon pairs with momenta $\mathbf{k}$ and $\mathbf{q}-\mathbf{k}$. Projecting the Hamiltonian onto these two sectors and retaining the cubic and quartic interactions~\cite{Zhang2026THED} gives
\begin{equation}\label{eq:thed_blocks}
  H(\mathbf{q})=
  \begin{pmatrix}
    H_{11}(\mathbf{q}) & H_{12}(\mathbf{q})\\
    H_{12}^{\dagger}(\mathbf{q}) & H_{22}(\mathbf{q})
  \end{pmatrix}.
\end{equation}
Because the polarized state has a single magnon band, $H_{11}(\mathbf{q})$ reduces to the LSWT energy $E(\mathbf{q})$ plus a static correction from normal ordering the quartic interaction in the Bogoliubov basis. In the two-magnon sector, $H_{22}(\mathbf{q})$ contains the pair energies and the quartic matrix elements that scatter one pair into another at fixed total momentum. The cubic interaction connects the two sectors, allowing a single magnon to split into a pair through $H_{12}^{\dagger}$ and the pair to recombine through $H_{12}$.

On a finite cluster of $N_s$ sites, THED diagonalizes Eq.~\eqref{eq:thed_blocks} to obtain the eigenstates $|n\rangle$ and their excitation energies $\omega_n$~\cite{Zhang2026THED,SunnyTHEDImplementation}. Each $\omega_n$ fixes the position of a spectral line. Its weight in a given component of the DSSF depends on how the corresponding spin operators connect the Bogoliubov magnon vacuum $|0\rangle$ to $|n\rangle$. Writing Eq.~\eqref{eq:dssf} in its Lehmann representation makes this dependence explicit,
\begin{equation}\label{eq:lehmann}
  S_{\alpha\beta}(\mathbf{q},\omega)=\frac{1}{N_s}\sum_n
  \langle 0|\big(S^\alpha_{\mathbf{q}}\big)^\dagger|n\rangle
  \langle n|S^\beta_{\mathbf{q}}|0\rangle\,
  \delta(\omega-\omega_n),
\end{equation}
where $S^\beta_{\mathbf{q}}=\sum_j e^{i\mathbf{q}\cdot\mathbf{r}_j}S^\beta_j$.

The connection to THL follows by eliminating the two-magnon sector from the resolvent of Eq.~\eqref{eq:thed_blocks}. This gives a Dyson equation for the one-magnon propagator in which propagation through the interacting magnon pair enters through a frequency-dependent self-energy,
\begin{equation}\label{eq:thl_dyson}
  \begin{split}
    G_1(\mathbf{q},z)
    &=\left[z-H_{11}(\mathbf{q})
    -\Sigma(\mathbf{q},z)\right]^{-1},\\
    \Sigma(\mathbf{q},z)
    &=H_{12}(\mathbf{q})
    \left[z-H_{22}(\mathbf{q})\right]^{-1}
    H_{12}^{\dagger}(\mathbf{q}),
  \end{split}
\end{equation}
where $z=\omega+i\eta$ with a small $\eta>0$. The three factors in $\Sigma$ describe the conversion of a single magnon into a pair, the propagation of that pair, and its recombination into a single magnon, read from right to left. Equation~\eqref{eq:thl_dyson} is exact within the projected space, where THL and THED therefore give the same one-magnon propagator.

The computational advantage of THL comes from how it evaluates the pair propagator $G_2(\mathbf{q},z)=[z-H_{22}(\mathbf{q})]^{-1}$. We write $H_{22}=X+V$, where $X$ contains the free-pair energies and $V$ scatters the two magnons while conserving Bogoliubov boson number. Repeated scattering by $V$ is summed to all orders by the two-magnon $T$ matrix,
\begin{equation}\label{eq:thl_ladder}
  G_2=G_2^{(0)}+G_2^{(0)}\,T\,G_2^{(0)},\qquad T=V+V\,G_2^{(0)}\,T,
\end{equation}
where $G_2^{(0)}(\mathbf{q},z)=[z-X(\mathbf{q})]^{-1}$ is the bare two-magnon propagator. The short range of the interaction allows us to express $V$ in a small set of real-space channels. The matrix to be inverted has one index per channel, so its dimension stays fixed as we refine the relative-momentum grid used for the Brillouin-zone integrals. We can therefore resolve the two-magnon continuum on a dense grid. For the calculations reported here, THL is more than an order of magnitude faster than THED.

An implementation of THL will be available in the v0.10 release of \texttt{Sunny.jl}~\cite{SunnyLadderImplementation} and was developed with AI assistance. This implementation includes contributions beyond the projected space from the anomalous static self-energy, the tadpole, and the cubic source vertex. These contributions have only a small effect on the spectra considered here (Appendix~\ref{app:thl_thed}).

The same resolvent gives direct access to the neutron response. Taking its matrix elements between the states created by the spin operators yields the positive-frequency part of the zero-temperature dynamical spin susceptibility,
\begin{equation}\label{eq:thl_chi}
  \chi_{\alpha\beta}(\mathbf{q},z)=\langle 0|\big(S^\alpha_{\mathbf{q}}\big)^\dagger
  \left[z-H(\mathbf{q})\right]^{-1}S^\beta_{\mathbf{q}}|0\rangle,
\end{equation}
from which the DSSF follows as
\begin{equation}\label{eq:thl_dssf}
  S_{\alpha\beta}(\mathbf{q},\omega)=\frac{1}{2\pi i N_s}
  \left[\chi_{\beta\alpha}(\mathbf{q},z)^*-\chi_{\alpha\beta}(\mathbf{q},z)\right].
\end{equation}
Block inversion of Eq.~\eqref{eq:thed_blocks} expresses the full projected resolvent in terms of $G_1$, $G_2$, and the coupling $H_{12}$ (Appendix~\ref{app:thl_dssf}).

Hybridization with the two-magnon continuum leaves signatures in both the line shape and the polarization dependence of the DSSF. At each momentum, the one-magnon LSWT spectrum consists of a single peak. Coupling to the continuum shifts this peak and transfers some of its spectral weight to the continuum. Where decay into a magnon pair is allowed, the peak also acquires an intrinsic width. Polarization provides a more direct signature of the mixing. For a field along $x$, the inelastic part of $S^x$ creates magnon pairs, while $S^y$ and $S^z$ create single magnons. A hybridized eigenstate can therefore be reached by both longitudinal and transverse spin operators, giving rise to the cross components $S_{xy}+S_{yx}$ and $S_{zx}+S_{xz}$, which vanish in LSWT (Appendix Fig. \ref{appfig:distinguishable_lswt}). In all spectra, we also apply a numerical Lorentzian broadening of half-width $\eta=0.2$ through the imaginary part of $z$. This broadening is present even for a stable magnon.

\subsection{Impact of the forward solver on Hamiltonian inference}
\label{sec:solver_impact}

\begin{figure*}[t]
\begin{center}
\setlabel{pos=nw,fontsize=\large,labelbox=false}
\includegraphics[width=\linewidth]{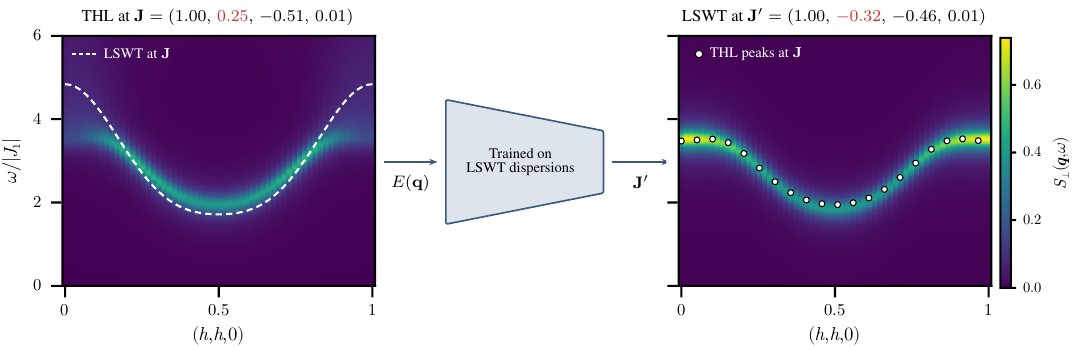}
\caption{\label{fig:main_result}
A network trained on LSWT, applied to an interacting spectrum. Left: THL spectrum $S_\perp(\mathbf{q},\omega)$ along $\mathbf{q}=(h,h,0)$ for $\mathbf{J}=(J_1,J_{zz},J_{\pm\pm},J_2)=(1.00,0.25,-0.51,0.01)$ and $J_{z\pm}=0$, at $B_x=B_\mathrm{sat}(\mathbf{J})+|J_1|$. The dashed line is the LSWT band at $\mathbf{J}$. Middle: the energies $E(\mathbf{q})$ of the main THL peak are the input of a Conv1D network trained on LSWT dispersions, which returns $\mathbf{J}'$. Right: LSWT spectrum at $\mathbf{J}'$, at $B_x=B_\mathrm{sat}(\mathbf{J}')+|J_1|$, with the THL peak energies (circles, every third momentum). LSWT at $\mathbf{J}'$ reproduces the THL peaks, although $J_{zz}$ (red) has the wrong sign. Both spectra have a Lorentzian half-width $\eta=0.2$ and share one color scale.
}
\end{center}
\end{figure*}

To illustrate the importance of choosing an appropriate solver when training a network, we directly examine whether a model trained on LSWT dispersions can recover the exchange couplings when applied to spectra where magnon interactions have shifted the measured peaks. Fig.~\ref{fig:main_result} illustrates the results of using a THL spectrum generated at known couplings $\mathbf{J}$ as a synthetic measurement. The input to the network is the energy of the main peak at each momentum. At the true couplings, the LSWT band misses the interaction-induced shifts of these peaks (left panel). At the couplings $\mathbf{J}'$ returned by the network, the LSWT band follows the peaks with a root-mean-square deviation of only $0.05|J_1|$ (right panel). This improvement comes at the expense of shifting other parameters unphysically, however: the inferred exchange anisotropy is $\Delta=J_{zz}/J_1=-0.32$, whereas the spectrum was generated with $\Delta=0.25$. A direct least-squares fit gives nearly the same anisotropy, $\Delta=-0.34$. Both procedures reproduce the peak positions by changing the exchange couplings to compensate for the magnon interactions omitted from LSWT.

A benchmark of 512 synthetic spectra in Appendix~\ref{app:lswt_thl} shows how often such an error can escape detection. For 174 spectra, the couplings returned by the LSWT-trained network reproduced the THL peak energies with a root-mean-square deviation below $0.1|J_1|$. Yet in 84 of those 174 apparently successful fits, at least one inferred coupling differed from its true value by more than $0.05|J_1|$. The limitations of LSWT were easier to recognize for another 243 spectra, for which even a direct least-squares search over the couplings left a deviation above $0.1|J_1|$, indicating that the spectra are unlikely to be described by any LSWT model whatsoever. In an experiment, such a residual would prompt scrutiny of the forward calculation before attributing the discrepancy to background, instrumental resolution, or an additional exchange interaction. The more difficult case is the one illustrated in Fig.~\ref{fig:main_result}, where the dispersion itself gives little reason to question the inferred Hamiltonian.

We may nonetheless resolve such ambiguous cases by extending consideration to features of the spectra beyond the bare dispersion. For the couplings $\mathbf{J}'$ in Fig.~\ref{fig:main_result}, the predicted peak intensities differ from the THL values by about $40\%$ of their mean, despite the close agreement in peak energies (Appendix~\ref{app:lswt_thl}). The incoherent two-magnon continuum supplies a further observable through its onset and the distribution of its weight in momentum and energy. In the remainder of the main text, we therefore train the inverse network on THL spectra, so that both the renormalized magnon peaks and the continuum enter the determination of the exchange couplings.

\section{Identifiability and Fisher Information}\label{sec:identifiable}

\begin{figure*}[t!]
\begin{center}
\includegraphics[width=0.99\linewidth]
{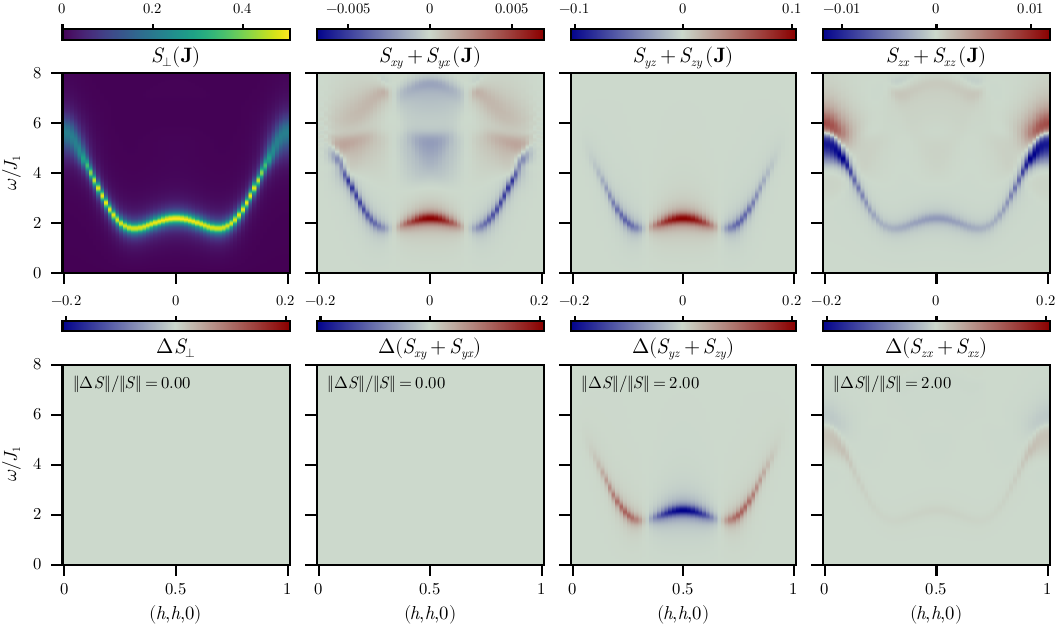}
\caption{\label{fig:distinguishable}
Measurement-dependent ambiguity in the sign of $J_{z\pm}$. The upper row shows the unpolarized response and selected DSSF tensor components along $\mathbf{q}=(h,h,0)$ for $\mathbf{J}=(J_1,J_{zz},J_{\pm\pm},J_{z\pm},J_2) =(1.0,0.7,0.2,0.2,0.051)$. The lower row shows $\Delta S=S(\mathbf{J}')-S(\mathbf{J})$ for $\mathbf{J}'=(1.0,0.7,0.2,-0.2,0.051)$, together with the relative norm $\|\Delta S\|/\|S(\mathbf{J})\|$ over the sampled $(\mathbf{q},\omega)$ grid. Both Hamiltonians are evaluated at the same field, $B_x=B_{\mathrm{sat}}(\mathbf{J})+|J_1|$, using the truncated-Hilbert-space ladder (THL) approach (Sec.~\ref{sec:thl}) with Lorentzian half-width $\eta=0.2$. Reversing $J_{z\pm}$ leaves all tensor components unchanged except $S_{yz}+S_{zy}$ and $S_{zx}+S_{xz}$. For these sign-reversing combinations, $\|\Delta S\|/\|S(\mathbf{J})\|=2$. Because $\widehat q_z=0$, neither combination contributes to the unpolarized response in Eq.~\eqref{eq:unpolarized_dssf}. %
}
\end{center}
\end{figure*}

In Sec.~\ref{sec:solver_impact}, neglecting magnon interactions led to the wrong sign of $J_{zz}$ even though the fitted dispersion closely reproduced the peak energies. Suppose now that the forward solver captures both the renormalized magnon dispersion and the two-magnon continuum. Does a proposed neutron measurement retain enough information to determine the exchange couplings uniquely?

Figure~\ref{fig:distinguishable} shows two Hamiltonians that differ only in the sign of $J_{z\pm}$ yet produce identical unpolarized spectra along $(h,h,0)$, including the continuum. The origin of this ambiguity is apparent in the DSSF tensor. 
Reversing $J_{z\pm}$ leaves the diagonal components and $S_{xy}+S_{yx}$ unchanged, but reverses the signs of $S_{xz}+S_{zx}$ and $S_{yz}+S_{zy}$. For in-plane momentum transfer $\widehat q_z=0$, so Eq.~\eqref{eq:unpolarized_dssf} projects out both sign-sensitive correlations. Sampling this cut more densely would therefore leave the sign undetermined. 

This example illustrates how the measurement protocol determines whether a coupling is identifiable. On a chosen parameter domain, the full Hamiltonian is identifiable when the forward map $\mathcal{F}$ assigns a distinct measured response to each parameter vector. Comparing the two spectra in Fig.~\ref{fig:distinguishable} exposes a discrete ambiguity in the sign of $J_{z\pm}$. We next use the Fisher information matrix to ask a complementary question: how sensitive is a proposed measurement to small changes in the couplings near a given Hamiltonian?

Let $S$ denote the vector of measured intensities on the sampled $(\mathbf{q},\omega)$ grid, and let $p(S|\mathbf{J})$ be its probability density for a Hamiltonian with couplings $\mathbf{J}$. The Fisher information matrix (FIM) quantifies how sensitively this distribution responds to small changes in the couplings~\cite{kay1993statistical}. It is defined by
\begin{equation}
  I(\mathbf{J})
  =
  \mathbb{E}\left[
  \nabla_{\mathbf{J}} \log p(S|\mathbf{J})\,
  \nabla_{\mathbf{J}} \log p(S|\mathbf{J})^T
  \right].
\end{equation}

The forward solver predicts the intensity vector $S(\mathbf{J})$. For Gaussian measurement noise with parameter-independent covariance $\Sigma$, the FIM depends on the spectral change produced by each coupling. These changes form the columns of the Jacobian $D$, with entries $D_{ak}=\partial S_a(\mathbf{J})/\partial J_k$, where $a$ labels a measured intensity. The FIM then takes the form derived in Appendix~\ref{app:fisherinfo},
\begin{equation}
  I(\mathbf{J})=D^T\Sigma^{-1}D.
  \label{eq:fisher_info_generic}
\end{equation}
If changes in two couplings produce nearly proportional changes in the measured intensities, one coupling can compensate for the other to first order. The measurement then constrains that combination of couplings only weakly.

To compare measurement protocols, we assume independent noise of unit variance for each intensity, so that $\Sigma_{ab}=\delta_{ab}$ and $I(\mathbf{J})=D^T D$. For full-tensor data, included to represent an upper bound on information that could be extracted from a polarized neutron scattering experiment, we use the three diagonal components $S_{\alpha\alpha}$ and the three symmetric combinations $(S_{\alpha\beta}+S_{\beta\alpha})/2=\mathrm{Re}\,S_{\alpha\beta}$ with $\alpha<\beta$. 

We approximate the Jacobian by central finite differences,
\begin{equation}
  \frac{\partial S(\mathbf{J})}{\partial J_k}\approx
  \frac{S(\mathbf{J}+\delta J\,\mathbf{e}_k)
       -S(\mathbf{J}-\delta J\,\mathbf{e}_k)}{2\delta J},
\end{equation}
where $\mathbf{e}_k$ is the unit vector along the parameter $J_k$
and $\delta J=10^{-3}$ in units of $|J_1|$.

The eigenvectors of the FIM identify combinations of exchange couplings to which the proposed measurement is more or less sensitive~\cite{liu2026observation}. For a normalized eigenvector $\mathbf{v}_k$, the relation $I=D^T D$ gives $\lambda_k=\|D\mathbf{v}_k\|^2$. A small eigenvalue therefore identifies a direction along which the couplings can vary together with little change in the measured spectrum. If all eigenvalues are positive, the smooth forward map is locally identifiable. A zero eigenvalue identifies a parameter direction along which the spectrum is unchanged to first order~\cite{rothenberg1971identification}.

We compare the most and least sensitive directions through the condition number
\begin{equation}
  \kappa = \frac{\lambda_{\max}}{\lambda_{\min}}.
\end{equation}
For the chosen parameter units and noise model, a large $\kappa$ indicates that the measurement constrains some combinations of couplings much more tightly than others. This comparison is local. In particular, a finite $\kappa$ at nonzero $J_{z\pm}$ is compatible with the exact ambiguity between opposite signs of $J_{z\pm}$ illustrated in Fig.~\ref{fig:distinguishable}.

\begin{table}[t!]
\caption{\label{tab:identifiable} Hamiltonian inference from THL spectra for different neutron-scattering protocols. The Fisher information matrix condition number $\kappa$ measures local conditioning at a reference Hamiltonian. We compute $\kappa$ using $N_k=60$ momentum points and reference couplings $(J_1,J_{zz},J_{\pm\pm},J_2)=(1,0.7,0.2,0.051)$. The four-parameter model fixes $J_{z\pm}=0$. The five-parameter model treats $J_{z\pm}$ as an unknown coupling, with reference value $J_{z\pm}=0.01$. Each momentum cut $(h,h,Q_z)$ runs from $(0,0,Q_z)$ to $(1,1,Q_z)$. The test mean squared error (MSE) averages the squared deviations of the inferred couplings from their true values over all inferred couplings and held-out Hamiltonians. For each protocol, we train the network of Sec.~\ref{sec:neuralnetworks} with hidden width $64$ on THL data with $N_k=12$, $N_E=20$, and $N_J=20$. The uncertainties on the MSE are standard deviations over $10$ training seeds. ``Full tensor'' denotes the three diagonal and three symmetric off-diagonal components of $S_{\alpha\beta}$ and represents an idealized upper bound on what could be measured in an experiment.}
\centering
\begin{ruledtabular}
\begin{tabular}{lcc}
Model / DSSF & MSE & $\kappa$ \\
\hline
\multicolumn{3}{c}{$4$ parameters ($J_{z\pm}$ excluded)} \\
\hline
Unpolarized, $(hh0)$
    & $1.54(26)\times10^{-4}$ & $1.19\times10^{2}$ \\ %
Full Tensor, $(hh0)$
    & $1.61(9)\times10^{-4}$ & $1.29\times10^{2}$ \\ %
Unpolarized, $(hh1)$
    & $1.43(17)\times 10^{-4}$ & $1.47\times10^{2}$ \\ %
Unpolarized, $(hh2)$
    & $1.43(34)\times10^{-4}$ & $1.47\times10^{2}$ \\ %
\hline
\multicolumn{3}{c}{$5$ parameters ($J_{z\pm}$ included)} \\
\hline
Unpolarized, $(hh0)$
    & $7.694(4)\times10^{-2}$ & $1.40\times10^{6}$ \\ %
Full Tensor, $(hh0)$
    & $9.35(73)\times10^{-5}$ & $2.51\times10^{3}$ \\ %
Unpolarized, $(hh1)$
    & $1.37(31)\times 10^{-5}$ & $1.31\times10^{4}$ \\ %
Unpolarized, $(hh2)$
    & $1.06(18)\times 10^{-5}$ & $4.90\times10^{3}$ \\ %
\end{tabular}
\end{ruledtabular}
\end{table}

When $J_{z\pm}$ is fixed to zero, the unpolarized in-plane scan recovers the four remaining couplings as accurately as the other measurement protocols tested. Table~\ref{tab:identifiable} compares the protocols using THL spectra. Across all four protocols, the FIM condition number $\kappa$ ranges from $119$ to $147$ at the reference Hamiltonian, and the mean squared error (MSE) of the inferred couplings stays near $1.5\times10^{-4}$ on held-out spectra [Eq.~\eqref{eq:mse} in Sec.~\ref{sec:neuralnetworks}]. The two quantities probe different aspects of inference. The condition number measures the relative sensitivity to different combinations of couplings near the reference Hamiltonian, while the test MSE measures the accuracy achieved by the network across the sampled parameter domain.

The choice of measurement becomes decisive when $J_{z\pm}$ is also unknown. The in-plane unpolarized spectrum is even in $J_{z\pm}$, so opposite signs give identical spectra (Fig.~\ref{fig:distinguishable}). This same symmetry makes the spectrum weakly sensitive to changes in $|J_{z\pm}|$ near zero. At the five-parameter reference point, $J_{z\pm}=0.01|J_1|$, the in-plane scan gives $\kappa=1.40\times10^6$. Across the test set, its MSE reaches $7.7\times10^{-2}$. Analysis of all tensor components of the spin-correlation function or an out-of-plane scan gives access to the sign-sensitive correlations, reducing $\kappa$ by factors of roughly $100$ to $560$ and the test MSE by three to four orders of magnitude.

The forward solver also affects local sensitivity. The LSWT benchmark in Table~\ref{apptab:app_lswt_identifiability} uses different parameter sets and momentum grids from those in Table~\ref{tab:identifiable}. For a direct comparison, Appendix~\ref{app:conditioning_jzpm} evaluates both solvers at the same couplings and momenta. For the in-plane unpolarized scan at the values of $|J_{z\pm}|$ tested between $0.05|J_1|$ and $0.2|J_1|$, including magnon interactions in THL lowers $\kappa$ by about a factor of three. The interacting spectrum thus improves local conditioning even in a geometry that leaves the exchange sign undetermined.

How can an experiment determine the sign of $J_{z\pm}$ with unpolarized neutrons? Moving the scan out of the plane gives the sign-sensitive correlations $S_{xz}+S_{zx}$ and $S_{yz}+S_{zy}$ finite weights in the measured intensity. For example, along a cut $(h,h,Q_z)$, these weights are $-\widehat q_x\widehat q_z$ and $-\widehat q_y\widehat q_z$, respectively [Eq.~\eqref{eq:unpolarized_dssf}]. Here, we choose $c/a = 5$, a representative ratio of the out-of-plane to in-plane lattice parameters in Yb-based delafossites~\cite{Xie2023CsYbSe2,Dai2021NaYbSe2,Scheie2024NaYbSe2,Scheie2024KYbSe2}, to illustrate our ideas.
In the five-parameter THL benchmark, moving the unpolarized scan from $(h,h,0)$ to $(h,h,1)$ lowers the test MSE from $7.7\times10^{-2}$ to $1.4\times10^{-5}$ (Table~\ref{tab:identifiable}). The LSWT benchmark shows the same trend (Table~\ref{apptab:app_lswt_identifiability}). The full-tensor in-plane scan gives the smallest condition number at the reference Hamiltonian, while the unpolarized $(h,h,1)$ and $(h,h,2)$ scans give the lowest inference errors across the test set. 

The analysis above can be extended in a few simple ways to assist in making decisions for a real experiment. The unit-variance choice used for $\kappa$ in Table~\ref{tab:identifiable} is a convenient benchmark that isolates the role the forward model plays in determining identifiability. It is straightforward, however, to extend this analysis to include a simple model of the measurement process, where the intensity in a given $(\mathbf q, \omega)$ bin is determined by a Poisson counting process with the rate of the process proportional to $S(\mathbf q,\omega)$ plus background and the intensity proportional to the number of counts generated by the process normalized by exposure time.  Specifically, if counts in bin $a$ of the $(\mathbf q,\omega)$ grid are distributed as $\mathrm{Poisson}(\lambda_a)$ with $\lambda_a=T(S_a+b_a)$, where $T$ is the exposure and $b_a$ the background, the Fisher information is exactly
\begin{equation}
I_{kl}(\mathbf J)=T\sum_a\frac{\partial_k S_a\,\partial_l S_a}{S_a+b_a},
\end{equation}
which is Eq.~\eqref{eq:fisher_info_generic} with $\Sigma=\mathrm{diag}[(S_a+b_a)/T]$ (Appendix~\ref{app:fisherinfo}). Relative to unit variance, this weighting raises the contribution of weak spectral regions, including the two-magnon continuum, by an amount set by the background. The condition number does not depend on $T$, and for the unpolarized cuts we examined it changes by less than a factor of 2.5 from its unit-variance value, so the $\kappa$ comparisons above do not depend on the noise model. The value of this formulation is that it allows one to frame uncertainties in terms of units of exposure time, which can serve as guidance when planning an experiment. The inverse of $I$ gives the Cram\'er--Rao bound \cite{Cramer1946, Rao1945}\cite[Ch.~3]{kay1993statistical} on the standard deviation of any unbiased estimate of each coupling, $\sigma_{\mathrm{CR},k}=[(I^{-1})_{kk}]^{1/2}$, which scales as $T^{-1/2}$ (Table~\ref{tab:poisson_fim}). These bounds give, for example, the exposure needed to determine a coupling to a given precision on a given cut. A discrete ambiguity such as the sign of $J_{z\pm}$ is not a local question, however, and requires a global criterion. We may measure it by the expected log-likelihood ratio between the two signs, a value which also grows linearly with exposure, as described in Appendix~\ref{app:fisherinfo}.

Regarding the selection of cut, the magnetic point group gives a clear indication of which cuts can carry the sign. With $\mathbf B\parallel\hat x$ along $\mathbf a_1$, the product of inversion, the twofold rotation about $\hat y$ and time reversal leaves the field invariant. For decoupled layers it maps every $\mathbf q$ in the plane normal to $\mathbf B$ onto itself and acts on the spin components as $(S_x,S_y,S_z)\to(r_xS_x,r_yS_y,r_zS_z)$ with $r=(1,-1,1)$, so
\begin{equation}
S_{\alpha\beta}(\mathbf q,\omega)=r_\alpha r_\beta\,\big[S_{\alpha\beta}(\mathbf q,\omega)\big]^*\qquad(\mathbf q\perp\hat x).
\label{eq:correlation_accessibility_selection_rule}
\end{equation}
Hence $\mathrm{Re}\,S_{yz}=\mathrm{Re}\,S_{xy}=0$ there, and the sign survives only in $S_{xz}+S_{zx}=2\,\mathrm{Re}\,S_{xz}$, which enters $S_\perp$ with weight $-\widehat q_x\widehat q_z$. It follows that no vertical-field measurement can detect the sign, even with polarized neutrons, while a cut in which $\mathbf q$ has components along both $\mathbf B$ and $\mathbf c$ can. 

While the cuts $(h,h,Q_z)$, with $Q_z\neq0$, satisfy this condition, they are not accessible in a real experiment: a cut can be measured only if the field is normal to its scattering plane (vertical-field magnet) or lies in it (horizontal-field magnet), and the $(h,h,Q_z)$ plane makes a $60^\circ$ angle with the field. The cut $(2h,-h,1)$ is accessible, however, in a horizontal-field magnet with $\mathbf c$ in the scattering plane. On this cut $\widehat q_y=0$, so the sign enters only through $S_{xz}+S_{zx}$. This correlation couples longitudinal and transverse fluctuations and vanishes without the cubic coupling between one- and two-magnon states (Appendix~\ref{app:thl_dssf}); LSWT therefore predicts an unpolarized intensity on $(2h,-h,L)$ that is even in $J_{z\pm}$. Along the idealized cuts $(h,h,Q_z)$, the one-magnon correlation $S_{yz}+S_{zy}$ also contributes, with weight $-\widehat q_y\widehat q_z$, which is why the LSWT benchmark resolves the sign there (Table~\ref{apptab:app_lswt_identifiability}). On the accessible cut, resolving the sign requires an interacting solver such as THL.
It should be cautioned, however, that resolving the sign on this cut takes roughly ten times the counts needed on $(h,h,1)$ (Table~\ref{tab:sign_counts}), which must be taken into account when requesting beam time. Details are given in Appendix~\ref{app:fisherinfo}. We emphasize these considerations to illustrate the value of studying synthetic spectra, computed with an appropriate solver, when planning an experiment: they determine which measurements can resolve a given model.

\section{Inferring exchange couplings from interacting spectra}\label{sec:neuralnetworks}

\begin{figure*}[t!]
\begin{center}
\includegraphics[width=0.99\linewidth]{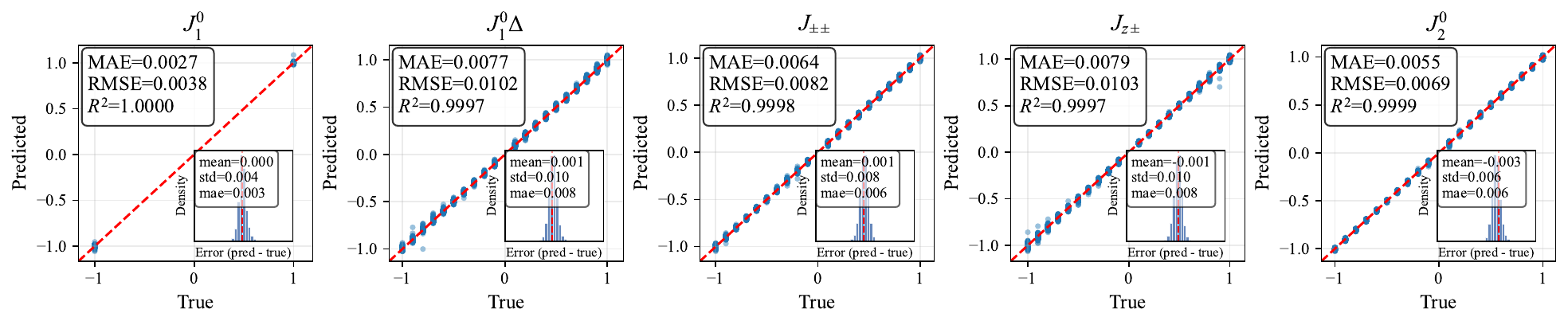}
\caption{\label{fig:error_plot} 
Parity plot comparing the predicted and true Hamiltonian parameters for a clean test set of $31{,}998$ samples in the five-parameter model with full-tensor data ($N_k=12,N_E=20, N_J=20$). The network is trained on $287{,}982$ samples. The dashed line denotes perfect prediction. The inset shows the distribution of parameter-prediction errors; the mean absolute relative error (MARE) is $1.81\%$.
}
\end{center}
\end{figure*}

Section~\ref{sec:identifiable} identified measurements that retain the sign of $J_{z\pm}$. How accurately can we recover all five exchange couplings from such a measurement? We use the THL solver of Sec.~\ref{sec:forward} to generate interacting spectra for Hamiltonians with known couplings. Each calculation provides a pair $(\boldsymbol{s}_i,\mathbf{J}_i)$, where $\boldsymbol{s}_i$ contains the DSSF data selected by the measurement protocol. From these pairs, we learn the parameters $\theta$ of a network that estimates the couplings from a spectrum,
\begin{equation}
  \widehat{\mathbf{J}}_i=f_\theta(\boldsymbol{s}_i).
\end{equation}
On the $31{,}998$ held-out Hamiltonians in the five-parameter benchmark, the network achieves a mean absolute relative error of $1.81\%$ using noise-free THL full-tensor spectra (Fig.~\ref{fig:error_plot}). In the four-parameter benchmark we fix $J_{z\pm}=0$ and infer the remaining couplings.

The input follows the spectrum along a chosen momentum path. At each momentum $k_j$, we represent the THL line shape by energy--intensity pairs,
\begin{equation}
  \boldsymbol{s}_i=\left\{E_n(k_j),\,I_n(k_j)\right\}_{n,j}.
\end{equation}
In the piecewise-linear representation, $E_n(k_j)$ is the energy of an adaptively selected knot and $I_n(k_j)$ is the spectral intensity there. These samples trace both the magnon peak and the distribution of weight through the two-magnon continuum. For the single-magnon LSWT spectra, the corresponding pair gives the band energy and its spectral weight. Appendix~\ref{app:adaptive_discretization} compares the adaptive knots with a Gaussian mixture representation of the continuum. The two give comparable inference errors, with slightly lower errors for knots on clean inputs and for Gaussian mixtures on noisy inputs in that benchmark. We store the features in an array of shape $(N_{\mathrm{ch}},N_k)$, with $N_{\mathrm{ch}}$ features per momentum and $N_k$ momenta. Fully connected networks and classical regression methods use the flattened array of length $N_{\mathrm{ch}}N_k$.

To generate the training and test data, we take $J_1=\pm1$ and sample each of $J_{zz}$, $J_{\pm\pm}$, and $J_2$ on a uniform grid of $N_J$ nonzero values in $[-1,1]$, in units of $|J_1|$. Varying $J_{z\pm}$ on the same grid increases the number of Hamiltonians from $2N_J^3$ to $2N_J^4$. We use $N_{J}=20$ for the THL benchmarks to limit the storage required for the discretized interacting spectra. For each Hamiltonian we set $B_x=B_{\mathrm{sat}}+|J_1|$, with $B_{\mathrm{sat}}$ the classical saturation field (Sec.~\ref{sec:model}). Interactions can shift the saturation field. For the five-parameter grid in Fig.~\ref{fig:error_plot}, we excluded the $20$ of $320{,}000$ parameter sets for which the THL calculation failed. Before optimization, we assign $90\%$ of the data to training and the remaining $10\%$ to testing. Unless stated otherwise, each parameter vector and its independently generated spectrum appear in only one set.

The networks in this section are trained on THL spectra. Appendix~\ref{app:lswt_thl} describes the LSWT-trained network used in Sec.~\ref{sec:solver_impact}, Appendix~\ref{app:lswt_results} reports the benchmarks trained and tested on LSWT spectra, and Appendix~\ref{app:lswt_convergence} examines their dependence on training-set size. We implement the networks in PyTorch~\cite{Paszke2019PyTorch,ketkar2021introduction} and train them on an NVIDIA RTX A6000 GPU. Generating the synthetic spectra accounts for most of the computational cost.

\subsection{Neural network architectures}\label{sec:architectures}

\begin{table}[t]
\caption{Four-parameter inversion of the unpolarized THL DSSF with four network architectures. Each input contains $N_k=60$ momenta and $N_E=20$ energy samples per momentum. The upper rows give the test errors for clean spectra and the lower rows give the errors for the Gaussian noise benchmark at $\sigma_\epsilon=0.1$. Parentheses give standard deviations over ten training seeds. Architecture details are given in Appendix~\ref{app:nn_architectures}.}
\label{tab:nn_arch_benchmark}
\centering
\small
\resizebox{\columnwidth}{!}{%
\begin{ruledtabular}
\begin{tabular}{lcccc}
Metric & FCNN & Conv1D & Transformer & ConvTransformer \\
\hline
MSE($10^{-4}$) & $4.76(12)$ & $\mathbf{1.54(12)}$ & $2.31(14)$ & $2.28(15)$ \\
MARE($\%$) & $3.7(1)$ & $\mathbf{2.04(8)}$ & $2.7(1)$ & $2.71(7)$ \\
\hline
MSE($10^{-4}$) & $21.3(20)$ & $\mathbf{7.48(35)}$ & $9.10(53)$ & $8.38(98)$ \\
MARE($\%$) & $8.2(4)$ & $\mathbf{4.98(8)}$ & $5.6(2)$ & $5.5(3)$ \\
\end{tabular}
\end{ruledtabular}
}
\end{table}

How much does the choice of network affect the inferred exchange couplings? Table~\ref{tab:nn_arch_benchmark} compares four architectures trained on the same four-parameter THL spectra. All four achieve mean absolute relative errors below $4\%$ on clean unpolarized data. The one-dimensional convolutional network (Conv1D) gives the lowest errors on both the clean and Gaussian noise benchmarks, $2.04\%$ and $4.98\%$, respectively. We use this architecture for the remaining benchmarks unless specified otherwise.

The architectures differ in how they use the momentum dependence of the DSSF. The fully connected neural network (FCNN) acts on the flattened spectral representation~\cite{goodfellow2016deep}. Conv1D first applies a convolution along the momentum axis, allowing it to follow how a peak or continuum feature changes between neighboring momenta, and then estimates the couplings with a fully connected regression head~\cite{kiranyaz20211d,malek2018one}. In our tests, additional convolutional or pooling layers gave no further improvement~\cite{wu2026geometric}. The transformer encoder uses attention to relate spectra at different positions along the cut~\cite{Vaswani2017}. The hybrid model applies a convolution before the transformer. We use a common hidden width of $64$, SiLU activations~\cite{Elfwing2018SiLU}, and linear outputs. The architectures have different numbers of trainable parameters (Appendix~\ref{app:nn_architectures}).

\subsection{Loss function and evaluation metrics}
\label{sec:metrics}

We train the network by minimizing the mean squared error in the exchange couplings. For $N$ Hamiltonians with $K$ inferred parameters each, the loss is
\begin{equation}\label{eq:mse}
  \mathcal{L}_{\mathrm{MSE}}(\theta)=\frac{1}{NK}\sum_{i=1}^{N}\sum_{k=1}^{K}
  \left(J_{i,k}-\widehat{J}_{i,k}\right)^2,
\end{equation}
where $J_{i,k}$ and $\widehat J_{i,k}$ are the true and inferred values of coupling $k$ for Hamiltonian $i$. We also report the mean absolute relative error (MARE),
\begin{equation}\label{eq:mare}
  \mathrm{MARE}=\frac{1}{NK}\sum_{i=1}^{N}\sum_{k=1}^{K}\frac{\left|J_{i,k}-\widehat{J}_{i,k}\right|}{|J_{i,k}|+\delta},
\end{equation}
where $\delta$ is a small positive constant. The MSE measures errors on the common energy scale, while the MARE gives greater weight to a given absolute error in a small coupling. We report both because relative errors can become large near a zero of an exchange coupling. Figure~\ref{fig:error_plot} resolves the five-parameter benchmark by coupling, including $J_{z\pm}$, whose sign is undetermined for the in-plane unpolarized scan (Sec.~\ref{sec:identifiable}).

\subsection{Noise augmentation and predictive uncertainty}
\label{sec:noise_augmentation}

How sensitive are the inferred couplings to uncertainty in the measured spectrum? For THL we perturb the energies and intensities of the discretized DSSF during training. Each feature $X_n(k)$ receives an independent Gaussian perturbation,
\begin{equation}
  \widetilde{X}_n(k) =X_n(k)+\epsilon_n(k)B_n^{(X)},
  \quad \epsilon_n(k)\sim\mathcal{N}(0,\sigma_\epsilon^2),
\end{equation}
where $B_n^{(X)}$ is the bandwidth for an energy channel and the maximum intensity for an intensity channel (Appendix~\ref{app:noise_model}). The amplitude $\sigma_\epsilon$ therefore specifies the perturbation relative to the scale of each feature. 

Noise augmentation discourages the network from changing its prediction sharply in response to small spectral perturbations~\cite{Bishop1995Noise,an1996effects}. For a scalar output and a prediction close to the target, a Taylor expansion gives the leading correction to the loss (Appendix~\ref{app:noise}),
\begin{equation}
\begin{aligned}
  \mathbb{E}_{\boldsymbol{\xi}}\!\left[\tfrac12
     (f_\theta(\boldsymbol{s}+\boldsymbol{\xi})-J)^2\right]
   &\simeq \tfrac12(f_\theta(\boldsymbol{s})-J)^2\\
   &\quad+\tfrac12\,(\nabla_{\boldsymbol{s}} f_\theta)^{T}
                C\,\nabla_{\boldsymbol{s}} f_\theta .
\end{aligned}
\end{equation}
Here $\boldsymbol{\xi}$ is the input perturbation and $C$ is its covariance. For the independent perturbations above, the diagonal entries of $C$ are $\sigma_\epsilon^2(B_n^{(X)})^2$. The correction penalizes sensitivity most strongly in the spectral features with the largest noise variance.

To estimate how uncertainty varies from one spectrum to another, we train a network to return a mean $\mu_{\theta,k}(\boldsymbol{s})$ and a variance $\sigma_{\theta,k}^2(\boldsymbol{s})$ for each coupling. The Gaussian negative log-likelihood is~\cite{KendallGal2017Uncertainty}
\begin{equation}
  \begin{split}
  &\mathcal{L}_{\mathrm{NLL}}(\theta)=\\
  &\frac{1}{NK} \sum_{i=1}^{N}\sum_{k=1}^{K}
  \left[
    \frac{\left(J_{i,k}-\mu_{\theta,k}(\boldsymbol{s}_i)\right)^2}{2\sigma_{\theta,k}^2(\boldsymbol{s}_i)}
    + \frac{1}{2} \ln\sigma_{\theta,k}^2(\boldsymbol{s}_i)
  \right],
  \end{split}
\end{equation}
The first term weights each squared error in a coupling by its predicted precision, and the logarithmic term penalizes large variances. The residual is evaluated in parameter space because the training Hamiltonians are known. Conventional fits to neutron data instead evaluate residuals in the measured spectrum~\cite{scheie2020multiphase}. Figure~\ref{fig:noise}(a) shows that the mean predicted variance increases with the noise amplitude used during training. These variances describe uncertainty conditional on the assumed Hamiltonian, forward solver, and noise model. The solver mismatch illustrated in Fig.~\ref{fig:main_result} can produce an additional systematic error.

To examine counting noise, we perform a separate LSWT benchmark in which counts are drawn from the broadened spectrum before the dispersion and intensity are extracted. For counts $C_{i\ell}\sim\mathrm{Poisson}[T(S_{i\ell}+b)]$, the background-subtracted intensity $\widetilde S_{i\ell}=C_{i\ell}/T-b$ has variance $(S_{i\ell}+b)/T$. The same count fluctuations thus affect both the extracted peak energy and its intensity. Appendices~\ref{app:poisson_noise} and~\ref{app:lswt_results} describe this benchmark and its inference errors.

\subsection{Generative decoder and self-consistency}
\label{sec:decoder}

\begin{figure}[t!]
\begin{center}
\includegraphics[width=0.95\linewidth]{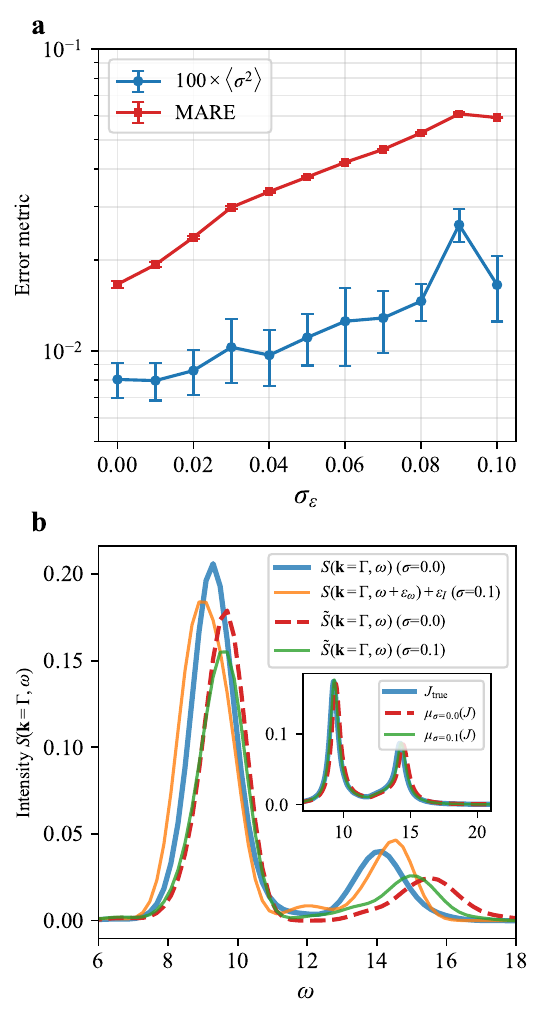}
\caption{\label{fig:noise} 
Effect of noise augmentation on the performance and predictive uncertainty of the inverse neural network. (a) The scaled mean predicted variance and the mean absolute relative error (MARE) of the inferred Hamiltonian parameters as functions of the noise amplitude $\sigma_\epsilon$ used during training. The test-error bars are standard deviation from 20 of the 200 training epochs. (b) Comparison of the DSSF reconstructed by the learned forward decoder with the reference THL result at $\mathbf{k}=\Gamma=(0,0)$ for the test sample $\mathbf{J}=(1,1,1,0,0.5)$. The inferred parameter vectors for training noise levels $\sigma_\epsilon=0$ and $\sigma_\epsilon=0.1$, with $J_{z\pm}=0$ fixed, are $\boldsymbol{\mu}_{\sigma=0}=(1.02(1), 1.05(2), 1.02(1), 0, 0.501(5))$ and $\boldsymbol{\mu}_{\sigma=0.1}=(1.00(1), 0.97(2), 1.00(1), 0, 0.509(8))$. The quoted uncertainties correspond to one standard deviation. The energy broadening used to reconstruct the intensity is $\eta=0.5$. The inset compares the direct THL spectra for the true parameter and the predicted mean parameters $\boldsymbol{\mu}_{\sigma}$ without extra Gaussian broadening after discretization $N_E=100$.
}
\end{center}
\end{figure}

Given a set of inferred couplings, a learned forward decoder $g_\phi$ provides a rapid estimate of the corresponding spectrum within the sampled parameter domain. We train it on the same type of Hamiltonian--spectrum pairs used for inversion. Its output is
\begin{equation}
  \widehat{\boldsymbol{s}}_i=g_\phi(\mathbf{J}_i),
\end{equation}
and its reconstruction loss averages over the $M$ spectral features,
\begin{equation}
  \mathcal{L}_{\mathrm{recon}}(\phi) = \frac{1}{NM}
  \sum_{i=1}^{N}\sum_{m=1}^{M} \left[s_{i,m}-g_{\phi,m}(\mathbf{J}_i)\right]^2.
\end{equation}
Passing the couplings inferred from a spectrum through the decoder gives the reconstructed input,
\begin{equation}
  \widehat{\boldsymbol{s}}_{\mathrm{recon}} = g_\phi\!\left(f_\theta(\boldsymbol{s})\right).
\end{equation}
The inverse network and the decoder can be trained separately with the parameter and reconstruction losses, respectively. In a joint training scheme, evaluating the reconstruction loss on $g_\phi(f_\theta(\boldsymbol{s}_i))$ makes it depend on the inferred couplings.

Figure~\ref{fig:noise}(b) compares the reconstructed spectrum at $\Gamma$ with a direct THL calculation for a test Hamiltonian. The decoder captures the dominant spectral features. The inset compares two physical calculations, one at the true couplings and one at the inferred couplings, and resolves changes in the multi-magnon response. This distinction matters experimentally. Agreement with a decoder checks consistency with its training solver. In Fig.~\ref{fig:main_result}, for example, an LSWT reconstruction of the dispersion would agree with the input peak energies even though the inferred $J_{zz}$ has the wrong sign. The decoder enables rapid exploration of candidate Hamiltonians. A direct interacting calculation at the inferred couplings then tests their predictions for both the peak and the continuum against the measured DSSF.

\subsection{Dependence on number of momentum points}
\label{sec:resolution}

\begin{table}[t]
\caption{Dependence of the four-parameter Conv1D inference error on the number of sampled momenta $N_k$ for unpolarized THL spectra. We retain $N_E=20$ energy knots per momentum. The upper rows give the errors for clean spectra and the lower rows give the errors for Gaussian noise of amplitude $\sigma_\epsilon=0.1$. Parentheses give standard deviations over ten training seeds.}
\label{tab:ne_length_noise}
\begin{ruledtabular}
{\fontsize{9.5}{12}\selectfont %
\begin{tabular}{lccccc}
$N_k$ & 12 & 20 & 30 & 60\\
\hline
MSE($\times10^{-4}$) & $1.36(11)$ & $1.20(9)$ & $1.22(7)$ & $1.34(7)$  \\
MARE($\%$)   & $1.84(8)$ & $1.80(5)$ & $1.83(8)$ & $1.95(6)$ \\
\hline
MSE($\times10^{-3}$) & $1.56(3)$ & $1.36(6)$ & $1.19(4)$ & $1.07(4)$  \\
MARE($\%$)   & $7.50(6)$ & $6.92(12)$ & $6.43(12)$ & $6.04(12)$ \\
\end{tabular}
}
\end{ruledtabular}
\end{table}

Once a momentum cut has been chosen, how densely must it be sampled? Table~\ref{tab:ne_length_noise} varies the number of sampled momenta from $N_k=12$ to $60$ at fixed $N_E=20$ energy knots per momentum. For the four-parameter model, the error on clean unpolarized THL spectra shows no systematic improvement over this range. The MARE stays between $1.80\%$ and $1.95\%$. With Gaussian noise of fixed amplitude $\sigma_\epsilon=0.1$ per feature, increasing $N_k$ lowers the error monotonically from $7.50\%$ to $6.04\%$. Additional momenta give the network more observations of spectral changes induced by the same couplings. The LSWT benchmark shows a similar dependence on noise (Appendix~\ref{app:lswt_resolution}). This result complements the geometry comparison in Sec.~\ref{sec:identifiable}. Denser sampling can improve precision for couplings to which a scan is already sensitive. Resolving the exact sign ambiguity of $J_{z\pm}$ requires access to a sign-sensitive correlation, for example by changing the momentum cut.

It is valuable to consider a similar analysis for the more experimentally realistic counting noise model, as introduced in Sec.~\ref{sec:identifiable} and discussed in Appendix~\ref{app:fisherinfo}. The benchmarks of Table~\ref{tab:ne_length_noise} perturb each feature with a fixed noise amplitude, so every additional momentum contributes an independent noisy observation. With counting statistics at a fixed exposure, finer momentum sampling divides the same counts among more bins, and the Poisson Fisher information of Sec.~\ref{sec:identifiable} is additive over disjoint bins (Appendix~\ref{app:fisherinfo}). The sum over momenta then approaches a Riemann sum,
\begin{equation}
I_{kl}\to T\int d\mathbf q\,d\omega\,\frac{\partial_kS\,\partial_lS}{S+b},
\end{equation}
which saturates once the sampling resolves the spectral structure. For the four-parameter model on $(h,h,0)$, the Cram\'er--Rao bounds at fixed exposure change by about 1\% between $N_k=12$ and 120, and by less than 0.1\% above $N_k=20$, whereas with a fixed variance per bin they fall as $N_k^{-1/2}$. In a counting measurement, finer momentum sampling therefore helps only insofar as the input representation retains more of the information already present in the counts; the bounds themselves fall as $T^{-1/2}$ with longer exposure.

\subsection{Classical regression baselines}
\label{sec:classical_baselines}

How much accuracy does the convolutional network gain over standard regression methods on the same interacting spectra? Table~\ref{tab:classical_baselines} compares linear, kernel, tree-based, and nearest-neighbor methods~\cite{hastie2009elements} using the flattened features of the clean four-parameter THL benchmark. Support-vector regression gives the lowest error among the classical methods, with an MSE of $6.80\times10^{-4}$ and a MARE of $3.01\%$. For the same sampling, $N_k=12$ and $N_E=20$, Conv1D achieves an MSE of $1.36\times10^{-4}$ and a MARE of $1.84\%$ (Table~\ref{tab:ne_length_noise}).

For the smoother LSWT spectra, Gaussian-process regression gives the lowest error in the corresponding benchmark (Appendix~\ref{app:lswt_classical_baselines}). For THL, Conv1D combines the lowest observed inference error with minibatch training on large sets of interacting spectra. Once trained, it estimates the couplings of each new spectrum without a new parameter search.

\begin{table*}[t!]
\caption{Classical regression baseline performance on unpolarized DSSF inversion (4 Hamiltonian parameters, $N_k=12, N_E=20$) under clean data using THL. Regression methods include Ridge, ElasticNet, Partial Least Squares (PLS), Kernel Ridge, Support Vector Regression (SVR), Gaussian Process Regression (GPR), Random Forest, Gradient Boosting, and k-Nearest Neighbor. The best performance is highlighted in bold.}
\label{tab:classical_baselines}
\centering
\scriptsize
\begin{ruledtabular}
\resizebox{\textwidth}{!}{%
\begin{tabular}{lccccccccc}
Metric & Ridge & ElasticNet & PLS & Kernel Ridge & SVR & GPR (cap 2000) & Random Forest &
Gradient Boosting & k-Nearest Neighbor \\
\hline
MSE     & $7.29\times10^{-3}$ & $1.29\times10^{-2}$ & $1.39\times10^{-2}$ & $4.16\times10^{-3}$ & $\mathbf{6.80\times10^{-4}}$ & $9.23\times10^{-4}$ & $2.28\times10^{-3}$ & $3.07\times10^{-3}$ &
$6.25\times10^{-3}$ \\
MARE  & $1.17\times10^{-1}$ & $1.61\times10^{-1}$ & $1.78\times10^{-1}$ & $8.18\times10^{-2}$ & $\mathbf{3.01\times10^{-2}}$ & $3.92\times10^{-2}$ & $6.68\times10^{-2}$ & $9.61\times10^{-2}$ & $1.57\times10^{-1}$ \\
\end{tabular}
}
\end{ruledtabular}
\end{table*}

\section{Practical workflow for Hamiltonian inference}
\label{sec:workflow}

Before beam time, the aim is to identify a measurement that distinguishes the Hamiltonians compatible with the crystal symmetry and with existing thermodynamic and spectroscopic data. For each candidate momentum cut and polarization channel, one can calculate the DSSF across this family of Hamiltonians and ask which spectral features change as the exchange couplings vary. In the field-polarized regime studied here, THL makes both the magnon peaks and the incoherent two-magnon continuum available for this comparison. To estimate which differences an experiment can resolve, the simulated spectra should incorporate the magnetic form factor, instrumental resolution, background, and expected counting statistics. With a common parameter scaling and an appropriate noise covariance, the Fisher information matrix identifies combinations of couplings to which each proposed scan is locally sensitive~\cite{liu2026observation,TeixeiraParente2023,Abuawwad2025}. Direct comparison of candidate spectra can also expose discrete ambiguities. In Sec.~\ref{sec:identifiable}, for example, opposite signs of $J_{z\pm}$ give identical unpolarized spectra along $(h,h,0)$, yet they can be distinguished along $(h,h,1)$ or $(2h,-h,1)$ using the same unpolarized technique. Training an inverse network for each candidate protocol and testing it on Hamiltonians withheld from training then gives a quantitative basis for choosing the scan: how accurately can it recover the coupling of interest at the expected noise level?

Once the spectrum has been measured, the network trained for the selected protocol provides an estimate of the couplings and, when trained with a noise model, their uncertainty under the assumed Hamiltonian family and forward solver. A learned forward decoder can quickly display the spectrum associated with those estimates. A more stringent test is to use the inferred couplings in a direct many-body calculation and predict a second momentum cut or polarization channel that was held out of the inference. If the prediction and measurement disagree, the origin of the discrepancy matters. As Sec.~\ref{sec:solver_impact} and Appendix~\ref{app:lswt_thl} illustrate, an approximate solver can mimic an interaction-induced peak shift by assigning an incorrect value to an exchange coupling. Checking the treatment of magnon interactions and the experimental response therefore precedes adding terms to the Hamiltonian. If several microscopic models remain viable, their predictions for the independent scan can guide the next measurement toward the region where they differ by more than the experimental uncertainty.

\section{Conclusion}\label{sec:conclusion}

Magnon interactions give a neutron spectrum additional ways to distinguish competing spin Hamiltonians. Two models with similar one-magnon dispersions can differ in the onset of the incoherent continuum, in the distribution of its spectral weight, and in the correlations between longitudinal and transverse spin fluctuations. The THL solver brings this information into Hamiltonian inference by treating the coupled one- and two-magnon sectors nonperturbatively in the thermodynamic limit. For the spectra considered here, THL was more than an order of magnitude faster than finite-cluster THED, making it practical to train an inverse network on interacting spectra. For the field-polarized triangular-lattice model, we generated interacting spectra for nearly $320{,}000$ Hamiltonians. A compact convolutional network trained on these spectra recovered the five exchange parameters with a mean absolute relative error of $1.8\%$ on held-out, noise-free data containing the full spin-correlation tensor.

An accurate fit to the magnon dispersion can nevertheless give a misleading account of the microscopic interactions. In our benchmark of 512 synthetic spectra, 174 predictions from a network trained on LSWT dispersions reproduced the interacting peak energies to within a root-mean-square deviation of $0.1|J_1|$. Nearly half of those predictions (84 of 174) placed at least one exchange coupling more than $0.05|J_1|$ from its true value. In the example of Fig.~\ref{fig:main_result}, the network and a direct least-squares fit both reversed the sign of $J_{zz}$. In each case, changing the exchange anisotropy allowed LSWT to mimic a peak shift caused by magnon interactions. THL predicts the interaction-induced peak shift and the accompanying continuum for each candidate Hamiltonian, so the inferred couplings can be tested against both.

The same simulations can determine which measurement will resolve a coupling of interest. When $J_{z\pm}$ was fixed to zero, an unpolarized in-plane scan sufficed to recover the four remaining couplings accurately. When $J_{z\pm}$ was allowed to vary, opposite signs produced identical unpolarized spectra along $(h,h,0)$. The sign information was present in the cross-correlations $S_{yz}+S_{zy}$, already present in LSWT, and $S_{xz}+S_{zx}$, generated by the mixing of one- and two-magnon states. Both can be accessed by moving the unpolarized scan out of the plane. Changing to an idealized cut of $(h,h,1)$ reduced the mean squared error across the five inferred couplings by a factor of more than $5{,}000$ on held-out synthetic spectra; the experimentally accessible cut, $(2h,-h,1)$, also allows the sign to be resolved, but at a penalty of requiring roughly ten times more counts than would be required along $(h, h, 1)$. The Fisher information matrix indicates which combinations of couplings a proposed measurement can constrain locally. Testing the inverse network on known Hamiltonians then shows how accurately those couplings can be recovered across the parameter range of interest. Both comparisons can be made before beam time is allocated.

Once trained, the inverse network makes the many-body calculation accessible to an experimentalist analyzing a new spectrum. It returns coupling estimates and, when trained with a noise model, uncertainties conditional on the assumed Hamiltonian family, solver, and measurement noise. A learned forward decoder provides an immediate preview of the spectrum at those couplings, while a direct THL calculation allows a more stringent test against an independent momentum cut or polarization channel. Applying this approach to a material will require incorporating instrumental resolution, magnetic form factors, background uncertainties, and counting statistics into the simulated interacting spectra. A Hamiltonian constrained by high-field measurements could then serve as input to many-body calculations of the lower-field phases. Active learning~\cite{TeixeiraParente2023} could also guide the experiment as it proceeds, directing the next scan toward the momenta and polarization channels where the Hamiltonians still consistent with the data make measurably different predictions.

Models are to theorists what materials are to experimentalists. The inverse scattering problem connects the two. It infers a material's microscopic Hamiltonian from its measured response, and the inferred Hamiltonian then predicts properties beyond those measured. Our results point to two complementary roles for artificial intelligence in solving this problem. 
The first is machine learning. A neural network trained on the output of an accurate forward solver turns many expensive many-body calculations into a fast inversion. 
The solver determines which physics the inference can capture, and neither component is sufficient on its own. 
The second is the development of the solvers themselves. AI assistants based on large language models draw on knowledge across many fields, so they can point to methods from outside a group's area of expertise and help adapt them to a specific many-body problem. 

The planned release of the ladder solver in \texttt{Sunny.jl}~\cite{SunnyLadderImplementation} will allow other groups to apply this analysis to their own spin Hamiltonians. 
With the appropriate probe operators, fast many-body solvers could also compute Raman, ESR, and RIXS spectra, as well as the optical conductivity.
This would enable multi-target inversion, in which a single Hamiltonian must account for several distinct responses, so that one probe can resolve models that another leaves ambiguous. 
More broadly, the same workflow applies to any inverse problem in which a forward model can generate training data across the relevant parameter range and quantify how sensitive candidate measurements are to those parameters.

\textit{Note added.} While preparing this manuscript, we became aware of Ref.~\cite{Paul2026DSFLearning}, which learns spin Hamiltonians from spectral moments of the DSSF without forward simulations. Our approach instead uses an interacting forward model to train neural networks on low-temperature spectra near saturation.

\section*{Acknowledgment}
We thank Jingyi Luo, Fletcher Williams, and Daniel Pajerowski for useful discussions.
Work at the University of Tennessee was supported by the National Science Foundation Materials Research Science and Engineering Center program through the UT Knoxville Center for Advanced Materials and Manufacturing (DMR-2309083). D.D. was supported by the U.S. Department of Energy (DOE), Office of Science, Basic Energy Sciences (BES) under the Genesis Mission BES AI Pathfinder Program. K.B. acknowledges support from the U.S. DOE, Office of Science, BES, Scientific User Facilities Division. H.Z. is funded in part by the Gordon and Betty Moore Foundation through Grant GBMF14251 to the American Chemical Society for the ACS-Moore Postdoctoral Bridge Award in Quantum Materials. 

\section*{Code availability}

The code accompanying this work is publicly available at \url{https://github.com/angkun-research/SpinDynamicsML}. The repository includes Julia scripts for spin-dynamics calculations and Python implementations of the inverse networks, regression benchmarks, uncertainty estimation, and learned forward decoder.

\appendix

\section{LSWT-trained inversion of THL spectra}
\label{app:lswt_thl}

This appendix details the test of Sec.~\ref{sec:solver_impact} and Fig.~\ref{fig:main_result}.
We draw 512 random parameter sets $\mathbf{J}$, with $J_1=\pm1$, the other couplings drawn uniformly from $[-1,1]$, and $J_{z\pm}=0$.
For each set, we compute the THL spectrum at $B_x=B_\mathrm{sat}(\mathbf{J})+|J_1|$ with a Lorentzian half-width $\eta=0.05$ and treat it as the measurement.
At each of the $N_k=60$ momenta along $(h,h,0)$, we extract the energy and the intensity of its main peak.
They are the input of two Conv1D networks trained on LSWT spectra at $|J_1|=1$, $J_{z\pm}=0$, and $B_x=B_\mathrm{sat}+|J_1|$.
One network takes only the peak energies, as in Fig.~\ref{fig:main_result}, and the other also takes the peak intensities.
Of the returned $J_1'$, we keep only the sign, so that $|J_1'|=1$.

Two criteria classify the outcome.
We call $\mathbf{J}'$ right if $J_{zz}'$, $J_{\pm\pm}'$, and $J_2'$ all lie within $0.05|J_1|$ of the true values.
The LSWT check compares the LSWT band at $\mathbf{J}'$, evaluated at $B_x=B_\mathrm{sat}(\mathbf{J}')+|J_1|$, with the THL peak energies.
Its misfit is the root-mean-square deviation over the 60 momenta, and $\mathbf{J}'$ passes the check if the misfit is below $0.1|J_1|$.
The two criteria give three cases. 
(i) $\mathbf{J}'$ is right and passes the check.
The many-body corrections are then small, and the inference succeeds.
(ii) $\mathbf{J}'$ fails the check, whether or not it is right.
The failure is visible, but reporting the parameters of the closest LSWT spectrum can give couplings far from $\mathbf{J}$.
This case also contains sets in which $\mathbf{J}'$ is right but fails the check, 38 of the 338 for the energies-only network, which a user cannot distinguish from a wrong answer.
(iii) $\mathbf{J}'$ is wrong, yet it passes the check.
The inference appears to succeed.
This is the most dangerous case, because no check within LSWT reveals the error, whereas in case (ii) the poor fit at least alerts the user.

As a control, we also fit the LSWT band directly to the peak energies by least squares.
This fit involves no network and gives the answer that a perfectly trained LSWT network would return.
We minimize the misfit with the Levenberg--Marquardt algorithm~\cite{Levenberg1944,Marquardt1963}  at $|J_1|=1$ for each sign of $J_1$, with the couplings restricted to the training range $[-1,1]$.
The minimization starts from 12 random points and from the true $\mathbf{J}$, so that the fit is at least as good as LSWT at $\mathbf{J}$.
For 243 of the 512 sets, even this fit leaves a misfit above $0.1|J_1|$, so that no LSWT parameters reproduce the THL peaks.

The example of Fig.~\ref{fig:main_result} belongs to case (iii).
Its couplings are $\mathbf{J}=(1.00,0.25,-0.51,0,0.01)$.
Magnon interactions lower the band near $\Gamma$ by about $1.4|J_1|$, where it meets the two-magnon continuum, and raise it near $h=1/2$.
The energies-only network returns $\mathbf{J}'=(1.00,-0.32,-0.46,0,0.01)$, and the misfit of LSWT at $\mathbf{J}'$ is $0.05|J_1|$.
The direct LSWT fit gives nearly the same couplings, $(1.00,-0.34,-0.48,0,0.00)$, with a misfit of $0.03|J_1|$.
The conclusion does not depend on the field at which LSWT is evaluated.
At the field of the measurement instead of $B_\mathrm{sat}(\mathbf{J}')+|J_1|$, the misfit of LSWT at $\mathbf{J}'$ is $0.06|J_1|$.
The peak intensities of LSWT at $\mathbf{J}'$ deviate from the THL ones by about $40\%$ of their mean, so a comparison of intensities could reveal the discrepancy.
The network that also takes the intensities as input returns, for this set, couplings that fail the LSWT check, with a misfit of $0.85|J_1|$.

\begin{table}[t]
\caption{Number of the 512 test sets in each case, for the two LSWT-trained networks applied to THL spectra.
$\mathbf{J}'$ is right if $J_{zz}'$, $J_{\pm\pm}'$, and $J_2'$ lie within $0.05|J_1|$ of $\mathbf{J}$, and it passes the check if the misfit of LSWT at $\mathbf{J}'$ is below $0.1|J_1|$.
For both networks, case (ii) contains the 243 sets for which no LSWT parameters pass the check.}
\label{apptab:lswt_thl_cases}
\centering
\begin{ruledtabular}
\begin{tabular}{lccc}
Network & (i) & (ii) & (iii) \\
\hline
Network, peak energies & 90 & 338 & 84 \\
Network, energies and intensities & 26 & 446 & 40 \\
\end{tabular}
\end{ruledtabular}
\end{table}

\begin{table}[t]
\caption{Number of the 512 test sets for which $J_{zz}'$, $J_{\pm\pm}'$, and $J_2'$ all lie within $0.05|J_1|$ or $0.1|J_1|$ of $\mathbf{J}$, for LSWT and THL spectra at the same $\mathbf{J}$.
The column headers give the tolerance in units of $|J_1|$.}
\label{apptab:lswt_thl_accuracy}
\centering
\begin{ruledtabular}
\begin{tabular}{lcccc}
& \multicolumn{2}{c}{LSWT input} & \multicolumn{2}{c}{THL input} \\
Network & $0.05$ & $0.1$ & $0.05$ & $0.1$ \\
\hline
Network, peak energies & 500 & 510 & 128 & 219 \\
Network, energies and intensities & 512 & 512 & 31 & 89 \\
\end{tabular}
\end{ruledtabular}
\end{table}

\begin{figure*}[t]
\centering
\includegraphics[width=\linewidth]{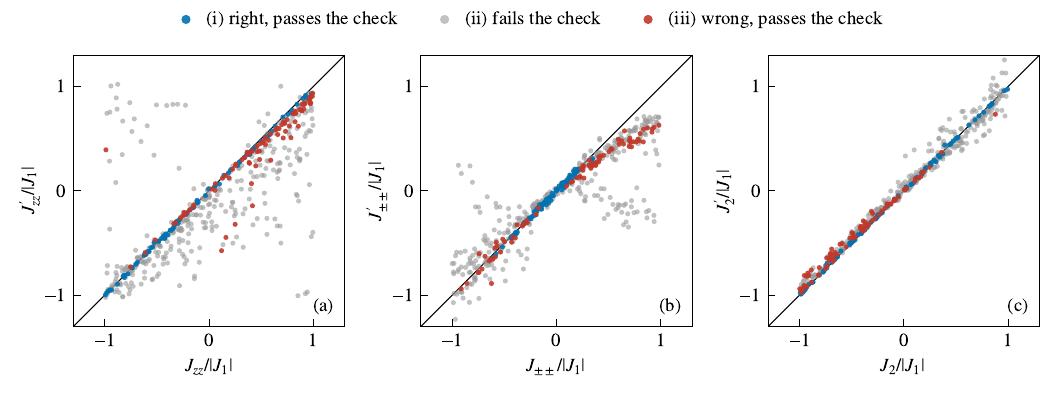}
\caption{\label{appfig:lswt_thl_parity}
Couplings $\mathbf{J}'$ returned by the network of Fig.~\ref{fig:main_result}, trained on LSWT dispersions, from the THL peak energies, against the true couplings $\mathbf{J}$, for the 512 test sets.
The colors mark the cases of Table~\ref{apptab:lswt_thl_cases}.
The solid line indicates $\mathbf{J}'=\mathbf{J}$.}
\end{figure*}

Table~\ref{apptab:lswt_thl_cases} lists the number of sets in each case.
The energies-only network returns wrong couplings that pass the check for 84 sets, the failure of Fig.~\ref{fig:main_result}.
It fails the check for 338 sets.
For 243 of them, no LSWT parameters pass the check, and a larger training set or a better network cannot change this.
For the other 95, an LSWT fit exists, but the network does not find it.
The network that also takes the intensities is more accurate on LSWT spectra (Table~\ref{apptab:lswt_thl_accuracy}).
The THL intensities, however, differ from the LSWT ones, and this network fails the check for 446 sets.

Table~\ref{apptab:lswt_thl_accuracy} shows that the networks themselves are accurate.
On LSWT spectra at the same 512 parameter sets, the energies-only network returns right couplings for 500 sets, and the network with intensities for all 512.
Nearly all other answers lie within $0.1|J_1|$ of $\mathbf{J}$.
On THL spectra, these numbers drop to 128 and 31.
The loss of accuracy thus comes from the forward model, not from the networks.

Figure~\ref{appfig:lswt_thl_parity} compares $\mathbf{J}'$ of the energies-only network with $\mathbf{J}$ for each coupling.
The errors are concentrated in $J_{zz}$ and $J_{\pm\pm}$, whereas $J_2$ lies within $0.1|J_1|$ of its true value in 452 of the 512 sets.
In case (iii), the root-mean-square errors of $J_{zz}$, $J_{\pm\pm}$, and $J_2$ are $0.22|J_1|$, $0.15|J_1|$, and $0.06|J_1|$, and the network underestimates $J_{\pm\pm}$ for $J_{\pm\pm}>0.5|J_1|$.
These errors are smaller than in case (ii), where the root-mean-square error of $J_{zz}$ is $0.50|J_1|$, but they pass the check.
In 5 of the 84 sets, among them the example of Fig.~\ref{fig:main_result}, $J_{zz}'$ has the wrong sign, with $|J_{zz}|>0.1|J_1|$.

\section{Spectral broadening and noise model}
\label{app:broadening}

LSWT produces a discrete set of excitation energies
$E_{n}(\mathbf{q})$ and corresponding spectral weights (intensities) $A_{n}(\mathbf{q})$. In contrast, the THL calculation is evaluated on a finite energy grid and produces a broadened intensity map in the $(\mathbf{q},\omega)$ plane. To compare the two descriptions on the same footing, we replace the delta functions in the LSWT dynamic structure factor by a normalized broadening kernel. For a discrete energy grid $\{\omega_\ell\}_{\ell=1}^{N_E}$, the broadened LSWT image at momentum $\mathbf{q}$ is therefore
\begin{equation}
S^{\mathrm{LSWT}}(\mathbf{q},\omega_\ell)
=
\sum_n A_n(\mathbf{q})\,
K_{\eta}\!\left[\omega_\ell-E_n(\mathbf{q})\right].
\end{equation}
The kernel $K_\eta(x)$ is a normalized function that replaces the delta function and has a characteristic width $\eta$.
We consider two common choices for the kernel: Lorentzian and Gaussian. A Lorentzian kernel is defined as 
\begin{equation}
K_{\eta}(x)
=
\frac{1}{\pi}\frac{\eta}{x^2+\eta^2},
\label{eq:lorentzian_kernel}
\end{equation}
where $\eta$ is the half-width at half-maximum. A Gaussian kernel can be used:
\begin{equation}
K_{\sigma}(x)
=
\frac{1}{\sqrt{2\pi}\sigma}
\exp\!\left(-\frac{x^2}{2\sigma^2}\right),
\label{eq:gaussian_kernel}
\end{equation}
where $\sigma$ is the standard deviation. This procedure converts the discrete LSWT bands into a smooth curve $S(\mathbf{q},\omega)$ for every momentum point and consequently into a two-dimensional heatmap. Without specific experimental information, we choose a Gaussian kernel with $\sigma=0.2$ for the broadening of LSWT data and noisy data.

To model uncertainty in the discrete data, noise can be introduced before the broadening operation. We use
\begin{equation}
  \begin{split}
    \widetilde{E}_n(\mathbf{q})
&=
E_n(\mathbf{q})+\delta E_n(\mathbf{q}),\\
\widetilde{A}_n(\mathbf{q})
&=
A_n(\mathbf{q})+\delta A_n(\mathbf{q}),
  \end{split}
\end{equation}
where the energy and intensity perturbations are sampled from the chosen noise model. The noisy broadened spectrum is then
\begin{equation}
\widetilde{S}(\mathbf{q}_i,\omega_\ell)
=
\sum_n \widetilde{A}_n(\mathbf{q}_i)\,
K_{\eta}\!\left[
\omega_\ell-\widetilde{E}_n(\mathbf{q}_i)
\right].
\label{eq:noisy_broadened_lswt}
\end{equation}
Thus, energy noise shifts the centers of the broadened peaks, while intensity noise changes their amplitudes. Applying the broadening after perturbing the discrete bands ensures that the noisy data have the same $(\mathbf{q},\omega)$ representation.

\subsection{Noise model for dispersion and intensity}
\label{app:noise_model}

To model experimental uncertainty, we perturb the dispersion and intensity representations
separately.
The perturbation amplitude is dimensionless and is controlled by a noise-level parameter $\sigma$.
For each sample and each spectral channel, independent Gaussian random variables
$\epsilon \sim\mathcal{N}(0,\sigma^2)$ are generated at every momentum or discretized spectral
index.
The noisy observable is then written as
\begin{equation}
  \widetilde{X}_n(k)= X_n(k) + \epsilon B_n^{(X)},
\end{equation}
where $B_n^{(X)}$ is a scale determined from the corresponding discretized index/band/intensity.
For a discretized index $E_n(k)$, we use its bandwidth,
\begin{equation}\label{eq:dispersion_noise_scale}
  B_n^{(E)}= \max_k E_n(k)-\min_k E_n(k).
\end{equation}
This choice makes the energy perturbation proportional to the variation of the band across momentum
space rather than to its absolute energy offset.
For an intensity band $S_n(k)$, or equivalently $I_n(k)$,
we instead use the maximum intensity as the scale,
\begin{equation}
  B_n^{(S)}=\max_k S_n(k).
\end{equation}
The dispersion and intensity perturbations therefore have different physical normalizations:
dispersion noise is measured relative to the bandwidth,
while intensity noise is measured relative to the largest intensity in the band.

For LSWT, the bandwidth in Eq.~\eqref{eq:dispersion_noise_scale} is evaluated from the band
dispersion $E_n(k)$ over the sampled momentum points.
For THL, the same construction is applied to the corresponding discretized spectral-index
representation; in this case, the maximum and minimum are taken over the discretized index used to
represent the band.
After these perturbations are applied, the noisy spectra are broadened and placed on the common
$(k,\omega)$ grid used by the regression model.

\subsection{Poisson counting noise}
\label{app:poisson_noise}

To model counting statistics, we also consider Poisson noise applied to the broadened DSSF,
which is more physically realistic for experimental measurements \cite{boothroyd2020principles}.
In this construction, fluctuations in the extracted dispersion and intensity arise from the same
noisy spectrum.
We describe the procedure for the single-band LSWT
representation, with excitation energy $E(\mathbf{q})$ and integrated spectral weight
$A(\mathbf{q})$.

We first broaden each excitation using the normalized Gaussian kernel in
Eq.~\eqref{eq:gaussian_kernel},
\begin{equation}
  S_{i\ell}
  = A(\mathbf{q}_i)
    K_{\sigma_{\mathrm{ins}}}
    \!\left[\omega_\ell-E(\mathbf{q}_i)\right],
\end{equation}
where $\sigma_{\mathrm{ins}}$ denotes the instrumental resolution width,
distinct from the noise amplitude in Appendix~\ref{app:noise_model}.
The implementation uses a full width at half maximum of $0.8$ in the energy units of the input data,
corresponding to $\sigma_{\mathrm{ins}}=0.8/\sqrt{8\ln 2}$.
For each sample, a common energy grid with spacing $\Delta\omega=0.01$ extends from the minimum
dispersion minus $6\sigma_{\mathrm{ins}}$ to the maximum dispersion plus $6\sigma_{\mathrm{ins}}$.

At each momentum and energy point, we independently sample the measured count,
\begin{equation}
  C_{i\ell}\sim
  \mathrm{Poisson}\!\left[T(S_{i\ell}+b)\right],
  \label{eq:poisson_counts}
\end{equation}
where $b$ is a uniform background in the same intensity normalization as $S_{i\ell}$,
and $T>0$ is an effective exposure factor converting intensity to expected counts.
The default background is $b=1$.
We subtract the known mean background after converting the counts back to intensity,
\begin{equation}
  \widetilde{S}_{i\ell}=\frac{C_{i\ell}}{T}-b.
\end{equation}
Consequently,
\begin{equation}
  \begin{split}
    \mathbb{E}[\widetilde{S}_{i\ell}] &= S_{i\ell},\\
    \operatorname{Var}(\widetilde{S}_{i\ell})
      &= \frac{S_{i\ell}+b}{T}.
  \end{split}
  \label{eq:poisson_intensity_moments}
\end{equation}
Subtracting the mean background therefore removes its average contribution but retains its counting
fluctuations.
The noise is intensity dependent, with a standard deviation proportional to $T^{-1/2}$.
Background-subtracted intensities
can be negative; these values are retained without clipping to avoid introducing a positive bias.

To recover the compact LSWT input representation, we calculate the zeroth and first moments of the
noisy spectrum,
\begin{equation}
  \begin{split}
    \widetilde{A}(\mathbf{q}_i)
      &= \Delta\omega\sum_\ell\widetilde{S}_{i\ell},\\
    \widetilde{E}(\mathbf{q}_i)
      &= \frac{\sum_\ell\omega_\ell\widetilde{S}_{i\ell}}
              {\sum_\ell\widetilde{S}_{i\ell}}.
  \end{split}
  \label{eq:poisson_lswt_moments}
\end{equation}
These expressions provide a moment-based estimate of the area and center of a single broadened peak.
They do not constitute a Poisson maximum-likelihood fit.
This approach mimics the single Gaussian-peak analysis (equivalent to $N=1$ Gaussian mixture model
as discussed in Appendix~\ref{app:gmm_discretization}). 

Figure~\ref{appfig:poisson_demo} shows an example of a Poisson-augmented LSWT spectrum.  Thus, Poisson augmentation generates dispersion and intensity perturbations jointly through counting fluctuations in the broadened spectrum. Unlike the Gaussian amplitude $\sigma_\epsilon$, which sets the standard deviation of each feature directly, $T$ sets the variance of each intensity bin, $(S_{i\ell}+b)/T$. The abscissa $10/T$ in Fig.~\ref{appfig:noise} is therefore proportional to a variance, the counting-noise standard deviation scales as $(10/T)^{1/2}$, and the Poisson and Gaussian curves should not be compared point by point.

\begin{figure}[h]
\centering
\includegraphics[width=\columnwidth]{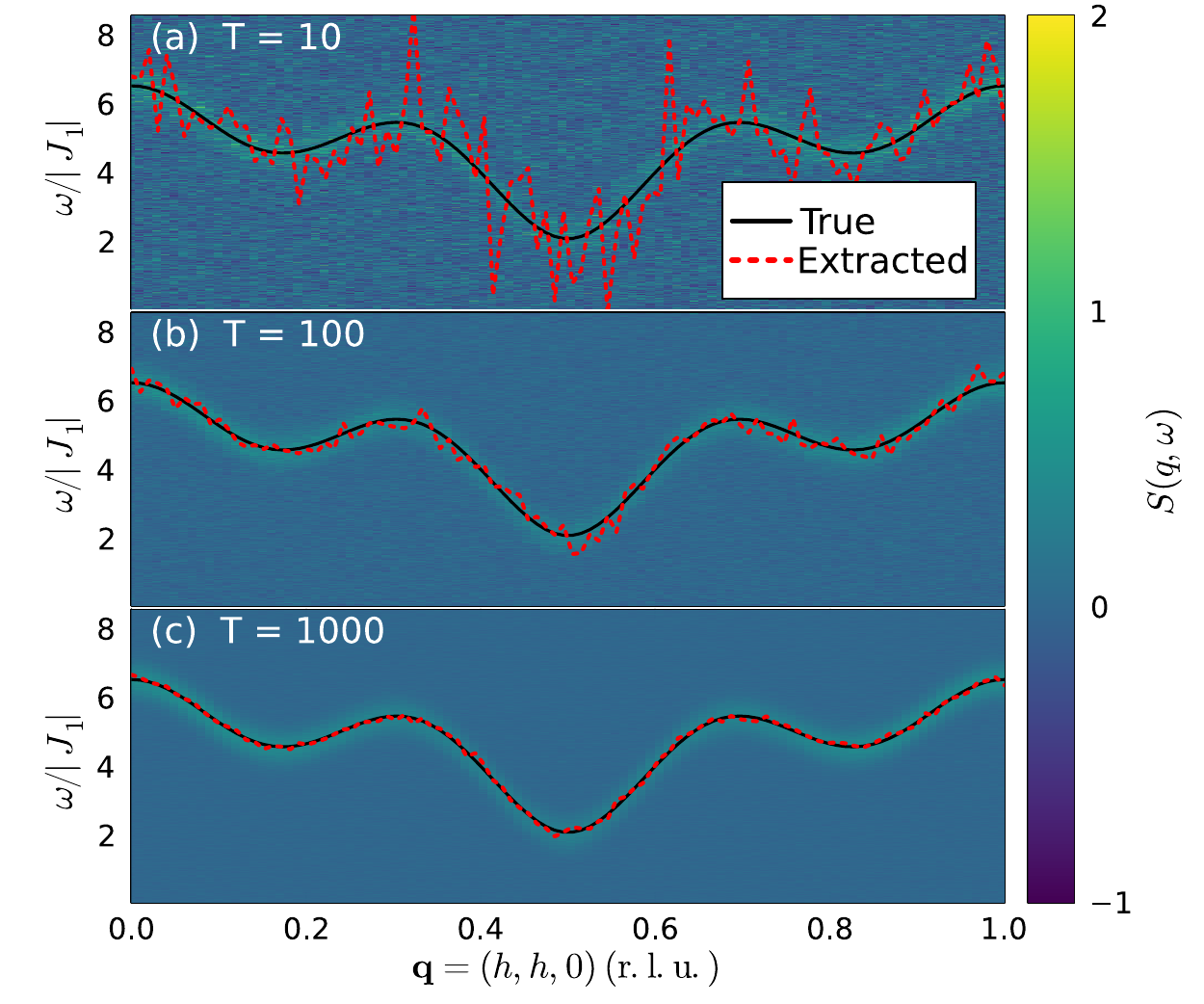}
\caption{Poisson counting noise in the Gaussian-broadened LSWT DSSF for effective exposure factors
(a) $T=10$, (b) $T=100$, and (c) $T=1000$.
The heatmaps show the background-subtracted noisy intensity $\widetilde{S}(\mathbf{q},\omega)$.
Black solid curves indicate the true LSWT dispersion,
while red dotted curves show the dispersion extracted from each noisy spectrum
[Eq.~\eqref{eq:poisson_lswt_moments}].
The Hamiltonian parameters are $\mathbf{J}=(1,-0.368,-1,0,0.474)$.
Increasing $T$ suppresses counting fluctuations and improves the recovery of the underlying
dispersion.}
\label{appfig:poisson_demo}
\end{figure}

For THL data, where we have continuum contributions instead of isolated LSWT peaks,
we need to consider discretization strategies (see Appendix~\ref{app:adaptive_discretization}).
The true spectrum may contain broad features (dense energy grid) and the adaptive discretization
should be applied after the Poisson augmentation.
However, storing the dense true spectrum is expensive.
If we apply the discretization both before and after the augmentation,
we risk losing important spectral details in the broad features.
We have not trained THL networks on Poisson counts, since this requires drawing counts from the dense spectrum before discretization; we leave it to future work. Appendix~\ref{app:fisherinfo} treats counting noise on the THL grid analytically through the Fisher information.
As a proof of concept, we only show the LSWT results for this Poisson noise in
Fig.~\ref{appfig:noise} (a).

\section{Dynamical structure factor in the THL method}
\label{app:thl_dssf}

Let $|0\rangle$ be the Bogoliubov vacuum, $\beta_{\mathbf{k}}|0\rangle=0$ for all $\mathbf{k}$.
At total momentum $\mathbf{q}$, the truncated space is spanned by the one-magnon state
$|\mathbf{q}\rangle=\beta^\dagger_{\mathbf{q}}|0\rangle$ and the two-magnon states
\begin{equation}\label{eq:thl_pair_states}
  |\mathbf{k},\mathbf{q}-\mathbf{k}\rangle
  =\frac{1}{\sqrt{1+\delta_{\mathbf{k},\mathbf{q}-\mathbf{k}}}}\,
  \beta^\dagger_{\mathbf{k}}\beta^\dagger_{\mathbf{q}-\mathbf{k}}|0\rangle,
\end{equation}
where each unordered pair $\{\mathbf{k},\mathbf{q}-\mathbf{k}\}$ is counted once.
These states are orthonormal.
The one-magnon amplitude $w_\beta$ and the two-magnon amplitudes $b_\beta(\mathbf{k})$ of a spin
component are the overlaps of $S^\beta_{\mathbf{q}}|0\rangle$ with these states,
\begin{equation}\label{eq:thl_amplitudes}
  w_\beta(\mathbf{q})=\langle\mathbf{q}|S^\beta_{\mathbf{q}}|0\rangle,\qquad
  b_\beta(\mathbf{k})=\langle\mathbf{k},\mathbf{q}-\mathbf{k}|S^\beta_{\mathbf{q}}|0\rangle,
\end{equation}
and $\mathbf{b}_\beta$ denotes the vector of the $b_\beta(\mathbf{k})$ over all pairs.
The projection of $S^\beta_{\mathbf{q}}|0\rangle$ onto the truncated space is therefore
$w_\beta|\mathbf{q}\rangle
+\sum_{\mathbf{k}}b_\beta(\mathbf{k})\,|\mathbf{k},\mathbf{q}-\mathbf{k}\rangle$.
Its zero-magnon part, $\langle 0|S^\beta_{\mathbf{q}}|0\rangle$, is nonzero only at
$\mathbf{q}=0$, contributes only at $\omega=0$, and is omitted.
The susceptibility of Eq.~\eqref{eq:thl_chi} is then
$\chi_{\alpha\beta}=(w_\alpha^*,\mathbf{b}_\alpha^\dagger)\,[z-H(\mathbf{q})]^{-1}\,
(w_\beta,\mathbf{b}_\beta)^{\mathsf T}$.
Inverting the $2\times2$ block matrix $z-H(\mathbf{q})$ of Eq.~\eqref{eq:thed_blocks} gives
\begin{equation}\label{eq:thl_block_inverse}
  \left[z-H\right]^{-1}=
  \begin{pmatrix}
    G_1 & G_1H_{12}G_2\\
    G_2H_{12}^\dagger G_1 & G_2+G_2H_{12}^\dagger G_1H_{12}G_2
  \end{pmatrix},
\end{equation}
with $G_1$ and $G_2$ of Eqs.~\eqref{eq:thl_dyson} and~\eqref{eq:thl_ladder}.
Hence,
\begin{equation}\label{eq:thl_chi_blocks}
  \begin{split}
    \chi_{\alpha\beta}
    &=w_\alpha^*\,G_1\,w_\beta
    +w_\alpha^*\,G_1 H_{12}G_2\,\mathbf{b}_\beta
    +\mathbf{b}_\alpha^\dagger\,G_2H_{12}^\dagger G_1\,w_\beta\\
    &\quad+\mathbf{b}_\alpha^\dagger\left(G_2
    +G_2H_{12}^\dagger G_1 H_{12}G_2\right)\mathbf{b}_\beta.
  \end{split}
\end{equation}

\begin{figure*}[t!]
\begin{center}
\includegraphics[width=0.9\linewidth]
{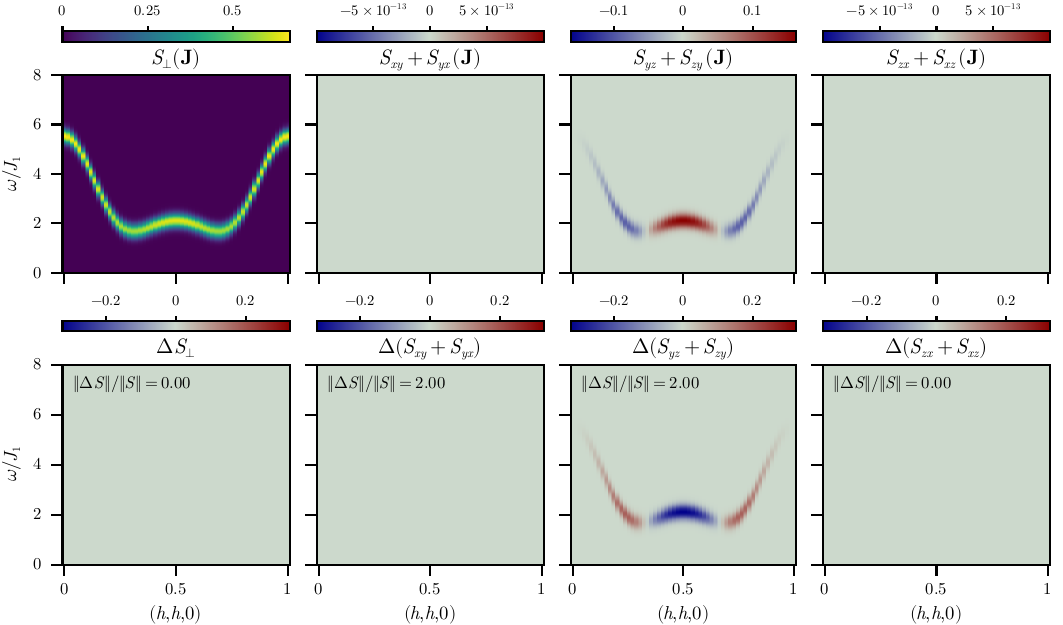}
\caption{\label{appfig:distinguishable_lswt}
Measurement-dependent ambiguity in the sign of $J_{z\pm}$ for the LSWT. The upper row shows the unpolarized response and selected DSSF tensor components along $\mathbf{q}=(h,h,0)$ for $\mathbf{J}=(J_1,J_{zz},J_{\pm\pm},J_{z\pm},J_2) =(1.0,0.7,0.2,0.2,0.051)$. The lower row shows $\Delta S=S(\mathbf{J}')-S(\mathbf{J})$ for $\mathbf{J}'=(1.0,0.7,0.2,-0.2,0.051)$, together with the relative norm $\|\Delta S\|/\|S(\mathbf{J})\|$ over the sampled $(\mathbf{q},\omega)$ grid. Both Hamiltonians are evaluated at the same field, $B_x=B_{\mathrm{sat}}(\mathbf{J})+|J_1|$, using the LSWT approach with Gaussian broadening $\sigma=0.2$.
}
\end{center}
\end{figure*}

For the field along $x$, the Holstein--Primakoff representation gives $S^x=S-b^\dagger b$, which is
even in the boson operators, whereas $S^y$ and $S^z$ are odd.
The Bogoliubov transformation is linear and preserves this parity.
Hence $w_x=0$ and $\mathbf{b}_y=\mathbf{b}_z=0$, and Eq.~\eqref{eq:thl_chi_blocks} reduces to
\begin{equation}\label{eq:thl_chi_parity}
  \begin{split}
    \chi_{\mu\nu}&=w_\mu^*\,G_1\,w_\nu,\qquad
    \chi_{x\mu}=\mathbf{b}_x^\dagger\,G_2H_{12}^\dagger G_1\,w_\mu,\\
    \chi_{\mu x}&=w_\mu^*\,G_1H_{12}G_2\,\mathbf{b}_x,\\
    \chi_{xx}&=\mathbf{b}_x^\dagger\left(G_2+G_2H_{12}^\dagger G_1H_{12}G_2\right)\mathbf{b}_x,
  \end{split}
\end{equation}
with $\mu,\nu\in\{y,z\}$.
The mixed components $\chi_{x\mu}$ and $\chi_{\mu x}$ are proportional to $H_{12}$.
The cross components $S_{xy}+S_{yx}$ and $S_{zx}+S_{xz}$ therefore vanish without the cubic
coupling, in particular in LSWT (Fig. \ref{appfig:distinguishable_lswt}).

\section{Comparison between THL and THED}
\label{app:thl_thed}

THL is formulated for the Nambu field
$\psi_{\mathbf{q}}=(\beta_{\mathbf{q}},\beta^\dagger_{-\mathbf{q}})^{\mathsf T}$, whose propagator
contains normal and anomalous parts~\cite{Beliaev1958a,Beliaev1958b}.
It obeys the Dyson equation
\begin{equation}\label{eq:thl_nambu}
  \begin{split}
    \mathbf{G}_1(\mathbf{q},z)={}&\big[z\tau_3-\mathcal{E}(\mathbf{q})
    -\boldsymbol{\Sigma}_\mathrm{st}(\mathbf{q})\\
    &+\delta\boldsymbol{\Sigma}_0(\mathbf{q})-\boldsymbol{\Sigma}(\mathbf{q},z)\big]^{-1},
  \end{split}
\end{equation}
where $z=\omega+i\eta$ as in Eq.~\eqref{eq:thl_dyson}, $\tau_3=\mathrm{diag}(1,-1)$, and
$\mathcal{E}(\mathbf{q})=\mathrm{diag}[E(\mathbf{q}),E(-\mathbf{q})]$ holds the LSWT energies.
Bold symbols denote $2\times2$ matrices in Nambu space.
The static self-energy $\boldsymbol{\Sigma}_\mathrm{st}$ comes from normal ordering the quartic
terms; its normal block, together with $E(\mathbf{q})$, forms $H_{11}(\mathbf{q})$.
The frequency-dependent self-energy $\boldsymbol{\Sigma}$ couples the magnon to the bath of
interacting pairs and is computed with the ladder of Eq.~\eqref{eq:thl_ladder}; its normal block
is $\Sigma(\mathbf{q},z)$ of Eq.~\eqref{eq:thl_dyson}.
The counterterm $\delta\boldsymbol{\Sigma}_0$ is described below.
Restricted to the one- and two-magnon space, the normal block of Eq.~\eqref{eq:thl_nambu} reduces to
$G_1$ of Eq.~\eqref{eq:thl_dyson}.

\begin{figure}[h]
\centering
\includegraphics{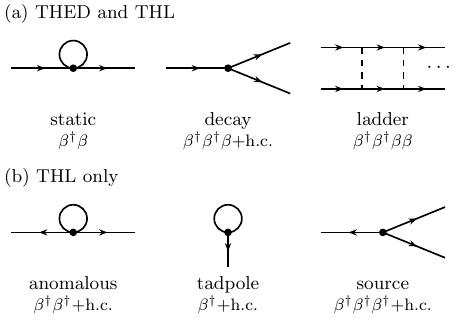}
\caption{\label{appfig:thl_diagrams}
Self-energy diagrams kept by (a) both THED and THL and (b) THL only.
Solid lines are Bogoliubov magnons, dots are interaction vertices, closed loops are contractions,
and dashed lines are the pair interaction.}
\end{figure}

Figure~\ref{appfig:thl_diagrams} shows the self-energy diagrams kept by the two methods.
Both methods keep the diagrams in Fig.~\ref{appfig:thl_diagrams}(a): the normal static self-energy
from normal ordering the quartic terms, the decay vertex that couples a magnon to a pair, and the
ladder of the pair interaction.
THL keeps in addition the diagrams in Fig.~\ref{appfig:thl_diagrams}(b), which lie outside the
one- and two-magnon space.
The anomalous static self-energy $\beta^\dagger\beta^\dagger+\mathrm{h.c.}$ couples a magnon to
three-magnon states.
The tadpole, a linear term from contracting two legs of the cubic vertex, tilts the spins away from
the field and corrects the static self-energy.
The source vertex $\beta^\dagger\beta^\dagger\beta^\dagger+\mathrm{h.c.}$ couples a magnon to pairs
through states with zero or four magnons.
The anomalous parts of these terms couple the magnon pole at $\omega$ to its mirror at $-\omega$,
and therefore shift the poles only at second order.

Beyond the self-energy, THL differs from THED in three further respects: the $1/S$ correction to
the single-magnon amplitudes of the spin operators, the static counterterm
$\delta\boldsymbol{\Sigma}_0$, which keeps Goldstone modes gapless in ordered phases, and the poles
of the retarded response at negative frequency.
They are described in the documentation of the implementation~\cite{SunnyLadderImplementation}.

\begin{figure*}[t]
\begin{center}
\includegraphics[width=0.99\linewidth]{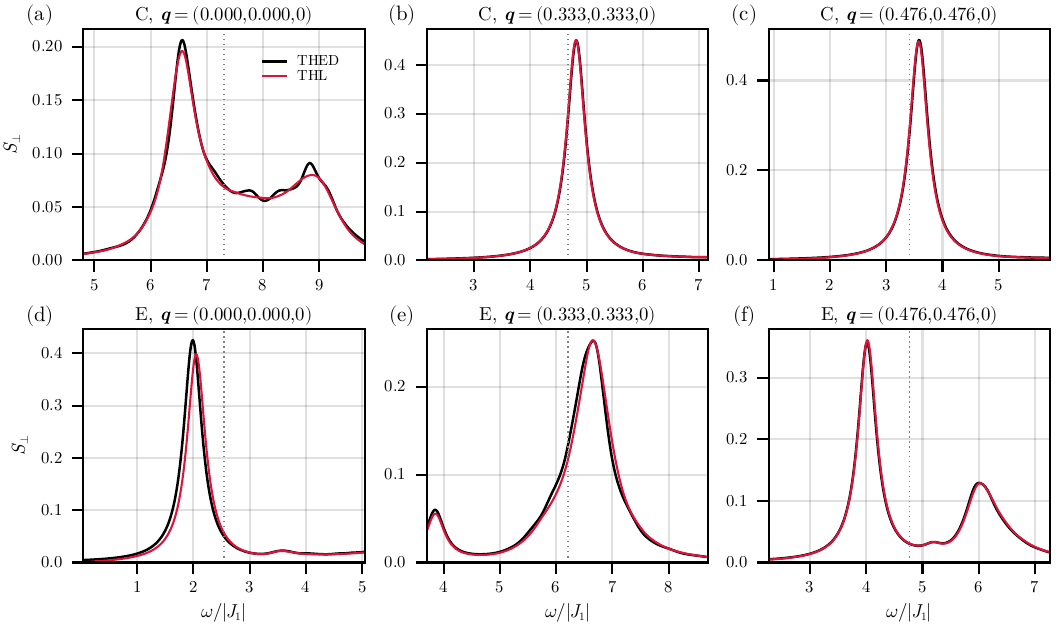}
\caption{\label{appfig:thl_thed}
Unpolarized DSSF $S_\perp$ at three momenta for models C [(a)--(c)] and E [(d)--(f)] of
Table~\ref{tab:thl_thed}, from THED on a $21\times21$ cluster (black) and from THL (red).
The dotted line marks the LSWT energy.
All spectra have a Lorentzian line shape of half-width $\eta=0.2$.
}
\end{center}
\end{figure*}

We compare the two methods for the five Hamiltonians of Table~\ref{tab:thl_thed}, at
$B_x=B_\mathrm{sat}+|J_1|$, along $\mathbf{q}=(h,h,0)$, and with $\eta=0.2$.
Model A has no cubic vertex, models B and C have weak and strong $U(1)$-breaking couplings, model D
is that of Fig.~\ref{fig:distinguishable}, and model E has a ferromagnetic $J_1$.
THED is evaluated on $15\times15$ and $21\times21$ clusters with 1000 Lanczos steps, and THL at the
same momenta.

\begin{table}[h!]
\caption{THL compared with THED on $15\times15$ and $21\times21$ clusters, at
$B_x=B_\mathrm{sat}+|J_1|$, along $\mathbf{q}=(h,h,0)$, with $\eta=0.2$, for
$\mathbf{J}=(J_1,J_{zz},J_{\pm\pm},J_{z\pm},J_2)$.
$\Delta S_\perp$ is the difference of the unpolarized DSSF over all $(\mathbf{q},\omega)$, in
percent of $\|S_\perp^\mathrm{THED}\|$.
$\Delta\omega_\mathrm{p}$ is the largest difference of the peak positions of $S_\perp$, in units of
$|J_1|$.}
\label{tab:thl_thed}
\centering
\small
\begin{ruledtabular}
\begin{tabular}{llcccc}
& & \multicolumn{2}{c}{$\Delta S_\perp$ (\%)} & \multicolumn{2}{c}{$\Delta\omega_\mathrm{p}$} \\
Model & $\mathbf{J}$ & $15\times15$ & $21\times21$ & $15\times15$ & $21\times21$ \\
\hline
A & (1,0.2,0,0,0.5) & 3.4 & 3.4 & 0.02 & 0.01 \\
B & (1,0.2,0.1,0.05,0.5) & 2.7 & 2.4 & 0.01 & 0.01 \\
C & (1,0.2,0.4,0.4,0.5) & 3.8 & 3.2 & 0.04 & 0.01 \\
D & (1,0.7,0.2,0.2,0.051) & 3.4 & 2.7 & 0.12 & 0.09 \\
E & (-1,-0.6,0.7,0.7,0.4) & 11.8 & 11.2 & 0.06 & 0.09 \\
\end{tabular}
\end{ruledtabular}
\end{table}

The two methods agree closely (Table~\ref{tab:thl_thed} and Fig.~\ref{appfig:thl_thed}).
For the antiferromagnetic models, $S_\perp$ differs by about $3\%$, and for the ferromagnetic
model E by about $11\%$.
The single-magnon peaks agree to within $0.12|J_1|$.
To trace these differences, we drop the additional terms from THL one at a time.
For the antiferromagnetic models, the anomalous static self-energy and the counterterm change
$S_\perp$ the most, by up to $4\%$ and $2\%$.
For model E, the counterterm dominates and changes $S_\perp$ by $16\%$.
The source vertex changes $S_\perp$ by less than $0.5\%$ for the antiferromagnetic models and by
$3\%$ for model E.

THL is also much cheaper.
On one CPU core, THED takes about $4\,\mathrm{s}$ per momentum on the $15\times15$ cluster and
about $14\,\mathrm{s}$ on the $21\times21$ cluster, whereas THL takes about $0.1\,\mathrm{s}$ for
all DSSF components.
Almost all of the THED time goes into building $H(\mathbf{q})$ as a dense matrix over the 113 and
221 pair states of the two clusters, so its cost grows roughly with the square of their number.
THL never forms this matrix.
It sums over the pairs of a grid of about $50\times50$ momenta once, and at each frequency it
inverts only a $27\times27$ matrix of interaction channels.

\section{Adaptive Discretization of the DSSF}
\label{app:adaptive_discretization}

For each fixed momentum point $\mathbf{q}$ (and for each measured DSSF channel),
we start from a dense spectrum $\{(\omega_i, S_i)\}_{i=1}^{n}$ and construct a compact
representation with at most $N$ support points
(knots).
The goal is to preserve sharp spectral features while reducing input size for regression.
Namely, given a fixed knot budget $N\ge 2$, the adaptive discretization greedily minimizes the
worst-case pointwise absolute interpolation error $\max_k |S_k-\hat S_k^{\mathrm{lin}}|$ by
iteratively adding the interior sample with the largest current residual.

The discretization is performed by iterative maximum-error refinement:
\begin{enumerate}
\item Initialize the kept index set with the two endpoints, $\{1,n\}$.
\item Treating neighboring kept points as line-segment boundaries, linearly interpolate $S(\omega)$
inside each segment.
\item Compute the absolute interpolation residual at every non-kept point,
\begin{equation}
e_k = \left|S_k - \hat S_k^{\mathrm{lin}}\right|.
\end{equation}
\item Find the point with the largest residual over all segments and add it to the kept set.
\item Repeat until $N$ knots are kept (or no interior point remains).
\end{enumerate}

This procedure is adaptive: smooth regions are represented with few points,
while high-curvature or peak regions receive more knots.
The final kept knots $(\omega_{\mathrm{keep}}, S_{\mathrm{keep}})$ define a piecewise-linear
approximation of the original DSSF spectrum and are used as the discretized input representation in
the inversion pipeline.

\begin{center}
\begin{minipage}{0.95\columnwidth}
\algotitle{Adaptive knot selection for DSSF discretization}
\label{alg:adaptive_knots}
\begin{algorithmic}[1]
\State \textbf{Input:} $\omega_1,\ldots,\omega_n$, $S_1,\ldots,S_n$, knot budget $N$
\State \textbf{Output:} $\mathcal{K}$ and $(\omega_{\mathcal{K}},S_{\mathcal{K}})$
\State $\mathcal{K}\gets\{1,n\}$
\While{$|\mathcal{K}|<N$}
  \State sort $\mathcal{K}$
  \State $\mathrm{bestErr}\gets -1,\ k^\star\gets -1$
  \For{$m=1$ to $|\mathcal{K}|-1$}
    \State $i\gets \mathcal{K}[m],\ j\gets \mathcal{K}[m+1]$
    \If{$j\le i+1$}
      \State \textbf{continue}
    \EndIf
    \For{$k=i+1$ to $j-1$}
      \State $\hat S\gets S_i+(S_j-S_i)(\omega_k-\omega_i)/(\omega_j-\omega_i)$
      \State $\mathrm{err}\gets |S_k-\hat S|$
      \If{$\mathrm{err}>\mathrm{bestErr}$}
        \State $\mathrm{bestErr}\gets\mathrm{err},\ k^\star\gets k$
      \EndIf
    \EndFor
  \EndFor
  \If{$k^\star<0$}
    \State \textbf{break}
  \EndIf
  \State $\mathcal{K}\gets\mathcal{K}\cup\{k^\star\}$
\EndWhile
\State \Return $\mathcal{K},\ \omega_{\mathcal{K}},\ S_{\mathcal{K}}$
\end{algorithmic}
\end{minipage}
\end{center}

\noindent
The computational cost is $O(Nn)$ in typical use, since each refinement pass scans points in current
segments.

\subsection*{Gaussian mixture model (GMM) discretization}
\label{app:gmm_discretization}

As an alternative to the linear knot selection, we also consider representing each spectrum
$\{(\omega_i,S_i)\}_{i=1}^n$ by a mixture of $N$ Gaussians \cite{bishop2006pattern}.
The DSSF is treated as an unnormalized density over $\omega$.
With $S_i^+=\max(S_i,0)$, $Z=\sum_i S_i^+$ and a uniform grid spacing $\Delta\omega$,
we define the normalized weights
\begin{equation}
p_i = \frac{S_i^+}{Z},
\end{equation}
which act as a probability mass function on the grid. We fit
\begin{equation}
\hat S(\omega)=\sum_{k=1}^{N} A_k\,\mathcal{N}(\omega;\mu_k,\sigma_k^2),
\qquad
A_k = \pi_k\, Z\,\Delta\omega,
\end{equation}
where $\mathcal{N}(\omega;\mu,\sigma^2)=(2\pi\sigma^2)^{-1/2}e^{-(\omega-\mu)^2/2\sigma^2}$,
and the mixture weights satisfy $\pi_k\ge0$ and $\sum_k\pi_k=1$.
The factor $Z\Delta\omega$ restores the overall scale of the spectrum,
so $A_k$ is the spectral weight carried by component $k$.
The compact representation is the set of $3N$ parameters $\{(\mu_k,\sigma_k,A_k)\}_{k=1}^N$.
Here $\mu_k$ gives the peak position, $\sigma_k$ its width, and $A_k$ its weight.
The parameters are fit by maximizing the $p$-weighted log-likelihood
\begin{equation}
\mathcal{L}=\sum_{i=1}^n p_i \log\!\Big(\sum_{k=1}^N
\pi_k\,\mathcal{N}(\omega_i;\mu_k,\sigma_k^2)\Big)
\end{equation}
We optimize this weighted likelihood using an adaptation of the standard expectation--maximization
(EM) algorithm \cite{Dempster1977EM,bishop2006pattern}, described below.

\paragraph{E-step.}
Given the current parameters, the responsibility of component $k$ for grid point $\omega_i$ is
\begin{equation}
r_{ik}=\frac{\pi_k\,\mathcal{N}(\omega_i;\mu_k,\sigma_k^2)}{\sum_{l}\pi_l\,\mathcal{N}(\omega_i;\mu_l,\sigma_l^2)}.
\end{equation}

\paragraph{M-step.} With $N_k=\sum_i p_i r_{ik}$, the parameters are updated as
\begin{align}
\pi_k &\leftarrow N_k, \\
\mu_k &\leftarrow \frac{1}{N_k}\sum_i p_i r_{ik}\,\omega_i, \\
\sigma_k^2 &\leftarrow \max\!\Big(\frac{1}{N_k}\sum_i p_i
r_{ik}(\omega_i-\mu_k)^2,\ \sigma_{\min}^2\Big),
\end{align}
followed by renormalization $\pi_k\leftarrow\pi_k/\sum_l\pi_l$.
The floor $\sigma_{\min}^2$ (we use $10^{-6}$) prevents components from collapsing onto a single
grid point.
Components with negligible mass ($N_k<10^{-12}$) are left unchanged.
We initialize $\mu_k$ at the $(k-\tfrac12)/N$ quantiles of $p$,
so components start in regions of high spectral weight.
We set $\sigma_k=(\omega_{\max}-\omega_{\min})/2N$ and $\pi_k=1/N$.
Iteration stops when $|\mathcal{L}^{(t+1)}-\mathcal{L}^{(t)}|<10^{-10}$ or after $200$ iterations.

\paragraph{Weight adaptation.}
Fixing $\pi_k=1/N$ (skipping the update of $\pi_k$) forces every component to carry equal spectral
weight.
Narrow components are then spent on weak features,
and a strong peak must be covered by several components.
Updating $\pi_k$ (weight adaptation) lets a single component with large $\pi_k$ describe a dominant
peak, while small-weight components describe weak or broad features.
This allocates the same budget $N$ more efficiently, as in the adaptive knot selection,
which places more knots where the spectrum varies rapidly.

\begin{figure}[h]
\centering
\includegraphics[width=\columnwidth]{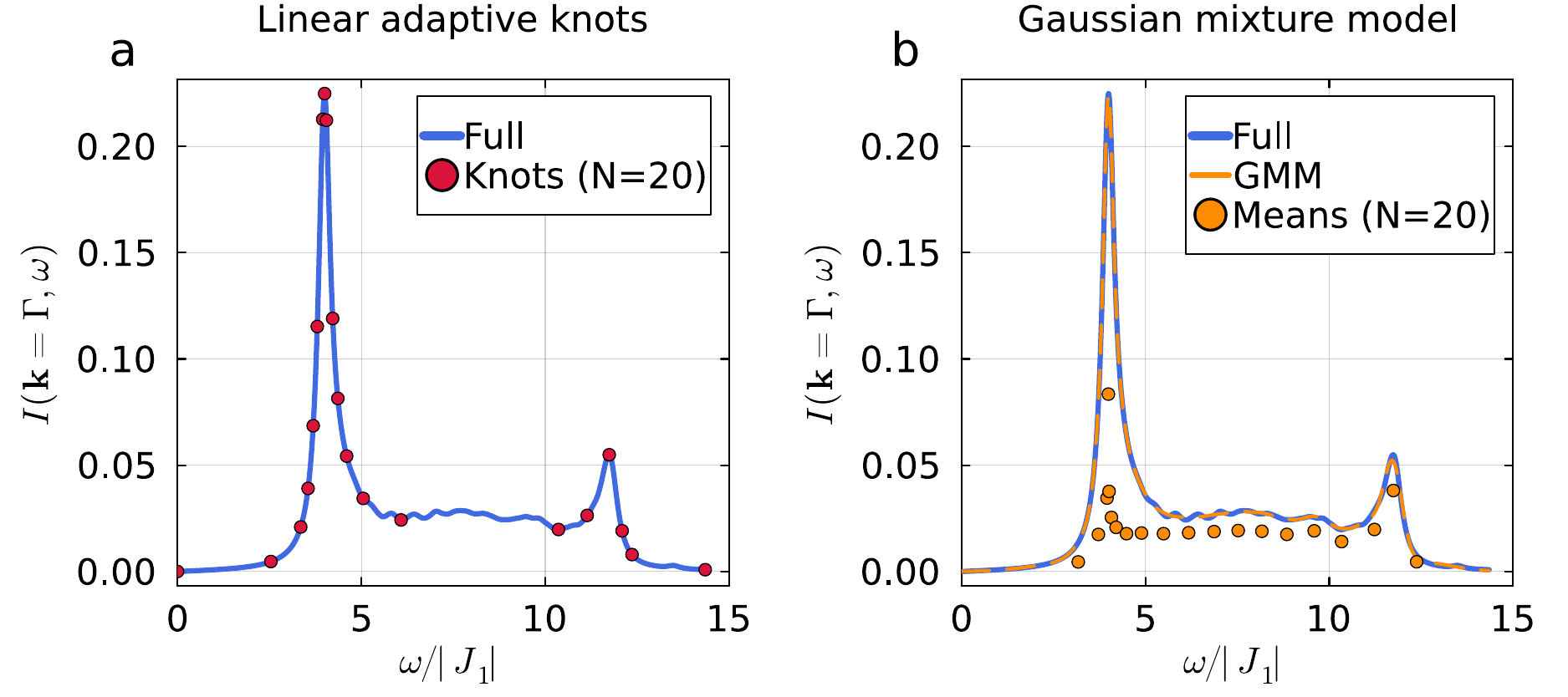}
\caption{Discretization of the DSSF at the $\Gamma$ point with $N=20$ parameters. Left: greedy piecewise-linear knots (red). Right: Gaussian-mixture reconstruction with adaptive weights (dashed orange), with component means marked at their peak heights $A_k/(\sigma_k\sqrt{2\pi})$. The full spectrum is shown in blue. The Hamiltonian parameters are $\mathbf{J}=(1.0,-0.368,-1.0,0,0.474)$.}
\label{fig:gmm_vs_knots}
\end{figure}

\noindent
Each EM iteration costs $O(Nn)$, so the total cost is $O(TNn)$.
Because EM maximizes a likelihood rather than a pointwise error,
it does not control $\max_k|S_k-\hat S_k|$ as the adaptive knot selection does.
It does give a smooth, parametric representation whose parameters have a direct physical
interpretation as peak position, width and weight.
Fig.~\ref{fig:gmm_vs_knots} compares the two discretizations at the $\Gamma$ point for $N=20$.

When it comes to the regression performance, the GMM discretization performs worse than the linear
discretization for clean data but better for noisy data,
as shown in Table~\ref{apptab:discretizations}.

\begin{table}[h!]
\caption{Neural network (Conv1D) performance on unpolarized DSSF with $4$ Hamiltonian parameters versus two different discretizations with clean data (first two rows) and noisy data with $\sigma=0.1$ (last two rows). The number of $k$-points is fixed at $N_k=60$ and the discretization energy points $N_E=20$ with THL solver. The results are from single seed with $100$ epochs.}
\label{apptab:discretizations}
\begin{ruledtabular}
\begin{tabular}{lccc}
Discretization & Linear & GMM\\
\hline
MSE & $1.40\times10^{-4}$ & $1.63\times10^{-4}$\\
MARE   & $2.05\times10^{-2}$ & $2.14\times10^{-2}$ \\
\hline
MSE & $1.07\times10^{-3}$  & $1.03\times10^{-3}$ \\
MARE   &  $6.14\times10^{-2}$ & $5.62\times10^{-2}$ \\
\end{tabular}
\end{ruledtabular}
\end{table}

\section{Fisher Information Matrix derivation} \label{app:fisherinfo}
\subsection{General Form}

The Fisher information matrix (FIM) \cite{kay1993statistical} is defined as
$$I(J) = \mathbb{E}\left[\nabla_J \log p(S|J) \nabla_J \log p(S|J)^T\right].$$
We assume that $\log p(S|J) = -\frac{1}{2} r(J)^T \Sigma^{-1}r(J) + \text{const}$ where the residual
is defined as $r(J) = S - S(J)$ and the covariance $\Sigma$ is parameter-independent.
The derivative is
\begin{align*}
\nabla_J \log p(S|J) &= \left( \frac{\partial r(J)}{\partial J} \right)^T \nabla_{r}
    \left( -\frac{1}{2} r(J)^T \Sigma^{-1} r(J) \right)\\
    &= [-\nabla_J S(J)]^T \Sigma^{-1}  [-r(J)],
\end{align*}
where $\frac{\partial r(J)}{\partial J}$ is  $-\nabla_J S(J)$,
since $S$ is the observed data and doesn't depend on $J$.
The outer product becomes
\begin{align*}
    &\nabla_J \log p(S|J) \nabla_J \log p(S|J)^T \\
    &=\left[ \nabla_J S(J)^T \Sigma^{-1} r(J) \right]
    \left[ \nabla_J S(J)^T \Sigma^{-1} r(J) \right]^T\\
    &= \nabla_J S(J)^T\Sigma^{-1}  r(J) r(J)^T \Sigma^{-1} \nabla_J S(J),
\end{align*}
where  $(AB)^T = B^T A^T$, and $(\Sigma^{-1})^T = \Sigma^{-1}$ as the covariance matrix is
symmetric.
Taking the expectation
\begin{align*}
    I(J) &= \mathbb{E}\left[ \nabla_J S(J)^T \Sigma^{-1} r(J) r(J)^T \Sigma^{-1} \nabla_J S(J)
    \right]\\
    &=\nabla_J S(J)^T \Sigma^{-1} \mathbb{E}\left[ r(J) r(J)^T  \right]\Sigma^{-1}  \nabla_J S(J)\\
    &=\nabla_J S(J)^T \Sigma^{-1}  \nabla_J S(J).
\end{align*}
Then, taking $\Sigma^{-1}=I$ gives $$I(J)=\nabla_J S(J)^T \nabla_J S(J)$$ as in the main text.

\subsection{With counting statistics}

For counting data, as is collected in a neutron scattering experiment, the likelihood is Poisson rather than Gaussian. With independent counts $C_a\sim\mathrm{Poisson}[\lambda_a(\mathbf J)]$, $\ln p(C|\mathbf J)=\sum_a[C_a\ln\lambda_a-\lambda_a]+\mathrm{const}$, so $\partial_k\ln p=\sum_a(C_a/\lambda_a-1)\,\partial_k\lambda_a$. Since $\mathbb E[(C_a-\lambda_a)(C_b-\lambda_b)]=\lambda_a\delta_{ab}$,
\begin{equation}
I_{kl}=\sum_a\frac{\partial_k\lambda_a\,\partial_l\lambda_a}{\lambda_a}=T\sum_a\frac{\partial_kS_a\,\partial_lS_a}{S_a+b_a}
\end{equation}
for $\lambda_a=T(S_a+b_a)$. This coincides with the Gaussian result with $\Sigma=\mathrm{diag}[(S_a+b_a)/T]$ and requires no Gaussian approximation. Splitting a bin over which the intensity varies slowly into two halves, each with mean $\lambda_a/2$ and derivative $\partial_k\lambda_a/2$, leaves its contribution unchanged, so at fixed exposure the Fisher information does not depend on the choice of binning. Any feature representation of the data, such as the knots of Appendix~\ref{app:adaptive_discretization}, is a processing of the binned counts, so its Fisher information cannot exceed that of the original binned data~\cite{Zamir1998}.

For any unbiased estimator $\hat{\mathbf J}$, $\mathrm{Cov}(\hat{\mathbf J})-I^{-1}$ is positive semidefinite~\cite{kay1993statistical}, so $\sigma_{\mathrm{CR},k}=[(I^{-1})_{kk}]^{1/2}$ is the smallest standard deviation any unbiased estimate of $J_k$ can have. Using the diagonal of the inverse, rather than $1/I_{kk}$, accounts for compensation by the other couplings. Because $I\propto T$, $\sigma_{\mathrm{CR}}\propto T^{-1/2}$.

A discrete ambiguity, such as the sign of $J_{z\pm}$ discussed in the text, cannot be resolved by local features. For two Hamiltonians $\mathbf J$ and $\mathbf J'$, the expected log-likelihood ratio in favor of $\mathbf J$ when $\mathbf J$ is true is the Kullback--Leibler divergence of their count distributions,
\begin{equation}
D(\mathbf J\,\|\,\mathbf J')=\sum_a\Big[\lambda_a\ln\frac{\lambda_a}{\lambda'_a}-\lambda_a+\lambda'_a\Big]\approx T\sum_a\frac{(S_a-S'_a)^2}{2(S_a+b_a)},
\end{equation}
which is exactly linear in $T$. This approximation holds when the spectra differ by little compared with $S+b$. For the sign of $J_{z\pm}$ we take $\mathbf J'$ with $-J_{z\pm}$ and minimize over its remaining couplings, $D_\mathrm{min}=\min_{\mathbf J'_\mathrm{rest}}D(\mathbf J\,\|\,(\mathbf J'_\mathrm{rest},-J_{z\pm}))$. By the Chernoff--Stein lemma, the error probability of the best test between the two decays exponentially in $D_\mathrm{min}$~\cite{CoverThomas2006}. We may take $D_\mathrm{min}=5$ as the threshold for a resolved sign. 

We evaluate these quantities on the $(\mathbf q,\omega)$ grid of $S_\perp$, with $N_k=60$ bin-centered momenta and an energy step of $\eta/4$, at the fixed field $B_0=B_\mathrm{sat}(\mathbf J_\mathrm{ref})+|J_1|$. Derivatives are central finite differences with $\delta J=10^{-3}$, and the Brillouin-zone quadrature grid of the THL calculation is held at its value for the reference couplings.
Refining the frozen grid by a factor of 1.5 changes $\kappa$ and $\sigma_{\mathrm{CR}}$ by less than 1\%. As expected from the rebinning argument above, varying the number of bin-centered momenta from $N_k=12$ to $120$ at fixed exposure changes $\sigma_{\mathrm{CR}}$ by about 1\% for the four-parameter model on $(h,h,0)$, and by less than 0.1\% above $N_k=20$. We minimize over $\mathbf J'_\mathrm{rest}$ with the Levenberg--Marquardt algorithm from several starting points. The background is $b=0.1\max S$, so that about 70\% of the counts are background; form factor and resolution are omitted. 
Tables~\ref{tab:poisson_fim} and~\ref{tab:sign_counts} summarize the results. For the four-parameter model, Poisson weighting changes the condition number by less than a factor of 2, so the conditioning in Table~\ref{tab:identifiable} does not depend on the noise model. The bounds translate directly into counts. On $(h,h,0)$, $N=10^6$ gives $\sigma_{\mathrm{CR}}\approx1.5\times10^{-3}|J_1|$ for $J_{zz}$, while a 10\% relative uncertainty in the small coupling $J_2$ needs only $N\approx7\times10^{3}$. The accessible cut $(2h,-h,1)$ constrains the four couplings about as well as $(h,h,0)$, whereas $(0,k,1)$ loosens the bounds on $J_1$ and $J_2$ by factors of about 3 and 2.4. Resolving the sign of $J_{z\pm}$ costs far more. The required counts scale as $J_{z\pm}^{-2}$ up to $|J_{z\pm}|\approx0.2$, are ten to fifteen times larger on $(2h,-h,1)$ than on the idealized $(h,h,1)$, and reach $4\times10^{6}$ at $J_{z\pm}=0.05$.

\begin{table}[t]
\caption{Fisher information under Poisson counting at the four-parameter reference couplings of Table~\ref{tab:identifiable}, $(J_1,J_{zz},J_{\pm\pm},J_2)=(1,0.7,0.2,0.051)$ with $J_{z\pm}=0$, at the fixed field $B_\mathrm{sat}+|J_1|$ of these couplings and with background $b=0.1\max S$. The unit-variance values, computed on the same grid with the field held fixed, lie within 10\% of Table~\ref{tab:identifiable}. The last four columns give the Cram\'er--Rao bounds on the standard deviations of the couplings as $\sigma_{\mathrm{CR}}\sqrt N$, in units of $|J_1|$, where $N$ is the total expected count on the cut, background included. The cut $(h,h,1)$ is an idealized geometry; $(2h,-h,1)$ requires a horizontal field containing $\mathbf c$, and $(0,k,1)$ a vertical field.}
\label{tab:poisson_fim}
\begin{ruledtabular}
\begin{tabular}{lcccccc}
Cut & $\kappa$, unit variance & $\kappa$, Poisson & $J_1$ & $J_{zz}$ & $J_{\pm\pm}$ & $J_2$\\
\hline
$(h,h,0)$ & 131 & 90 & 0.48 & 1.47 & 0.91 & 0.44\\
$(h,h,1)$ & 161 & 104 & 0.50 & 1.57 & 0.95 & 0.45\\
$(2h,-h,1)$ & 131 & 87 & 0.53 & 1.49 & 0.66 & 0.49\\
$(0,k,1)$ & 220 & 171 & 1.58 & 1.08 & 0.93 & 1.07\\
\end{tabular}
\end{ruledtabular}
\end{table}

\begin{table}[t]
\caption{Total expected counts $N$ on the cut, background included, needed for $D_\mathrm{min}=5$, at $(J_1,J_{zz},J_{\pm\pm},J_2)=(1,0.7,0.2,0.051)$ with $J_{z\pm}$ varied; $J_{z\pm}=0.2$ is the Hamiltonian of Fig.~\ref{fig:distinguishable}. The background is $b=0.1\max S$. On $(h,h,0)$ and $(0,k,1)$, $D_\mathrm{min}=0$ to numerical precision, and the sign cannot be resolved at any exposure. The cut $(h,h,1)$ is an idealized cut and $(2h,-h,1)$ an experimentally accessible one.}
\label{tab:sign_counts}
\begin{ruledtabular}
\begin{tabular}{lcc}
$J_{z\pm}$ & $(h,h,1)$ & $(2h,-h,1)$\\
\hline
0.05 & $4.1\times10^{5}$ & $4.3\times10^{6}$\\
0.1 & $1.0\times10^{5}$ & $1.1\times10^{6}$\\
0.2 & $2.6\times10^{4}$ & $2.9\times10^{5}$\\
0.5 & $5.1\times10^{3}$ & $7.5\times10^{4}$\\
\end{tabular}
\end{ruledtabular}
\end{table}

\section{Regularization of the regression from noise\label{app:noise}}

In this section, we provide a detailed derivation of the regularization effect of training a neural
network with noisy data \cite{Bishop1995Noise,an1996effects}.
For convenience, we denote a single sample of the DSSF data as $X$ and the corresponding Hamiltonian
parameter as $y$ (assume a single parameter for simplicity).
The neural network is denoted as $f(X)$, which maps the DSSF data to the predicted Hamiltonian
parameter.
We start with the loss function for clean data:
\begin{equation}
  \begin{split}
    \mathcal{L}_\mathrm{clean}(X) &= \frac{1}{2} (f(X) - y)^2.
  \end{split}
\end{equation}
Consider adding Gaussian noise $\epsilon \sim \mathcal{N}(0,\sigma_\epsilon^2)$ to the input data
$X$, resulting in a noisy input $X + \epsilon$.
The loss function for the noisy data is then
\begin{equation}
  \begin{split}
    \mathcal{L}_\mathrm{noisy}(X) &= \frac{1}{2} (f(X + \epsilon) - y)^2.
  \end{split}
\end{equation}
Assuming that the noise is small, the Taylor expansion of $f(X + \epsilon)$ around $X$ gives
\begin{equation}
  \begin{split}
    f(X + \epsilon) &\approx f(X) + \nabla_X f(X) \cdot \epsilon
    + \frac{1}{2} \epsilon^T H(X) \epsilon,
  \end{split}
\end{equation}
where $[H(X)]_{ij} = \left(\frac{\partial^2 f}{\partial X_i \partial X_j}\right)$ is the Hessian
matrix of second derivatives of $f$ with respect to $X$.
Substituting this expansion into the noisy loss function and keeping only terms up to order
$O(\Vert{}\epsilon\Vert{}^2)$ yields:
\begin{widetext}
  \begin{equation}   
    \begin{split}
      \frac{1}{2} \left( f(X + \epsilon) - y \right)^2 &\approx \frac{1}{2} \left( f(X)
      - y \right)^2 + \underbrace{\left( f(X) - y \right) \epsilon^T \nabla_X f(X)}_{\text{Linear in
      }\epsilon} + \underbrace{\frac{1}{2} \left( \epsilon^T \nabla_X f(X)
      \right)^2}_{\text{Quadratic in }\epsilon} + \underbrace{\frac{1}{2} \left( f(X)
      - y \right) \epsilon^T H(X) \epsilon}_{\text{Quadratic in }\epsilon}\\
      \mathbb{E}_{\epsilon} \left[ \frac{1}{2} (f(X + \epsilon) - y)^2 \right] &\approx \frac{1}{2}
      \left( f(X) - y \right)^2 + \frac{1}{2} \sigma_\epsilon^2 \Vert \nabla_X f(X) \Vert^2
      + \frac{1}{2} \left( f(X) - y \right) \sigma_\epsilon^2 \mathrm{Tr}[H(X)],
    \end{split}
\end{equation}
\end{widetext}
where we have used the fact that $\mathbb{E}[\epsilon] = 0$ and
$\mathbb{E}[\epsilon \epsilon^T] = \sigma_\epsilon^2 I$.
Thus, assuming that the model has converged and the prediction is close to the true value, i.e.,
$f(X) \approx y$, we can neglect the last term and obtain the following approximation for the
expected loss function with noisy data:
$$ \mathcal{L}_\mathrm{noisy}(X) \approx \mathcal{L}_\mathrm{clean}(X) + \frac{1}{2}
\sigma_\epsilon^2 \Vert \nabla_X f(X) \Vert^2.
$$

\section{Neural Network Architectures}\label{app:nn_architectures}

\begin{figure*}[t!]
\begin{center}
\includegraphics[width=1.0\linewidth]{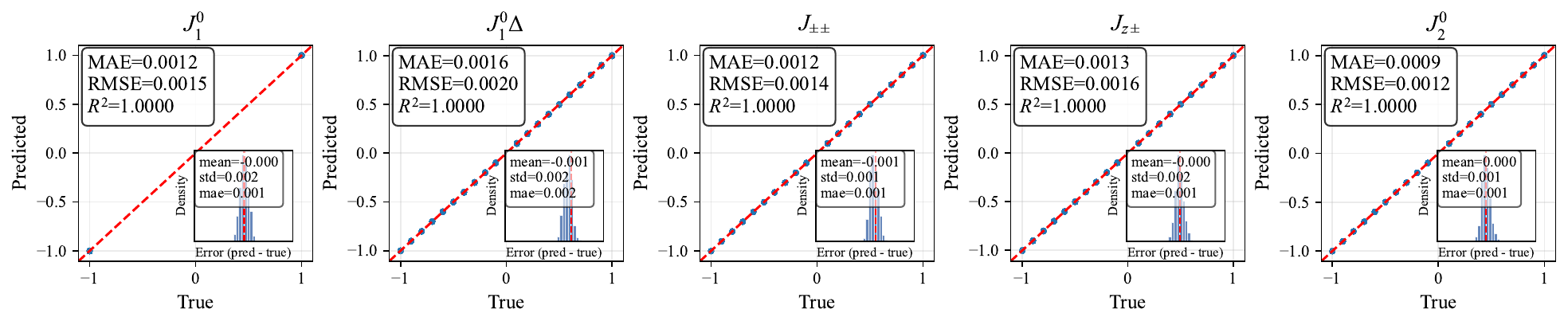}
\caption{\label{appfig:lswt_error_plot} 
The parity plot of the predicted Hamiltonian parameters versus the true Hamiltonian parameters for a test set of $32,000$ samples with $5$ Hamiltonian parameters and tensor data under LSWT solver ($N_k=N_J=20$). The dashed line indicates perfect prediction and there is no noise augmentation. Inset: the error distribution of the predicted Hamiltonian parameters. The mean absolute relative error (MARE) is $0.30\%$.
}
\end{center}
\end{figure*}

We provide a detailed description of the neural network architectures used in our experiments, which follows the spirit of minimalist design principles. This appendix summarizes the four regression models used in the benchmark: FCNN, Conv1D, Transformer, and ConvTransformer. All models take the same input tensor and predict the same Hamiltonian-parameter vector, so the comparison is architecture-only.

For a batch of $B$ spectra, the input tensor has shape $(B,N_{\mathrm{ch}},N_k)$, where $N_k$ is the number of sampled momenta and $N_{\mathrm{ch}}$ is the number of spectral features at each momentum. For a single DSSF component represented by $N_E$ knots, these features are the knot energies and intensities, so $N_{\mathrm{ch}}=2N_E$. The regression output contains one estimate for each varied exchange coupling, giving four outputs when $J_{z\pm}=0$ is fixed and five when it is varied.

Unless stated otherwise, the models use a hidden width of $64$ and SiLU activations. The final regression layer is linear and returns unconstrained coupling estimates.

\begin{enumerate}
  \item \textbf{Fully Connected Neural Network (FCNN).}

  This is the simplest baseline.
  The model first flattens the full DSSF input into one vector per sample.
  It then applies a small MLP:
  (flatten) $\rightarrow$ (linear to hidden width) $\rightarrow$ SiLU $\rightarrow$ (linear to
  output dimension).

  Note that this architecture is the most minimal FCNN that can perform the regression task. 

  \item \textbf{1D Convolutional Network (Conv1D).}

  This model adds one convolution stage before regression:
  (Conv1D with circular padding) $\rightarrow$ SiLU $\rightarrow$ (flatten) $\rightarrow$ (linear to
  hidden width) $\rightarrow$ SiLU $\rightarrow$ (linear to output).

  Circular padding is used so boundary points along the spectral axis are treated periodically.
  In practice, this architecture captures short-range patterns in neighboring bins before global
  regression.
  More convolution layers and pooling were tested, but they did not improve performance,
  so we report the simplest version.

  \item \textbf{Transformer Encoder Network (Transformer).}

  Here the spectral axis is treated as a sequence.
  The input is transposed to sequence-first form, then projected to hidden width:
  (transpose) $\rightarrow$ (linear input projection) $\rightarrow$ (single Transformer encoder
  layer) $\rightarrow$ (flatten) $\rightarrow$ (MLP head).

  The encoder layer uses multi-head self-attention (4 heads) with feedforward width equal to twice
  the hidden width.
  This architecture is designed to capture long-range dependencies across distant spectral
  locations.

  \item \textbf{Hybrid Convolution-Transformer Network (ConvTransformer).}

  This model combines local feature extraction and global attention:
  (Conv1D with circular padding) $\rightarrow$ SiLU $\rightarrow$ (transpose to sequence form)
  $\rightarrow$ (single Transformer encoder layer) $\rightarrow$ (flatten) $\rightarrow$ (MLP head).

  Intuitively, the convolution front-end builds local descriptors,
  and the transformer stage then models interactions between those descriptors across the full
  spectrum.
\end{enumerate}

Practical comparison goal:
- FCNN tests whether a minimal global regressor is sufficient.
- Conv1D tests whether local spectral structure improves inversion.
- Transformer tests whether long-range spectral coupling is important.
- ConvTransformer tests whether combining local and global mechanisms yields additional gains.

\section{Additional Results for Linear Spin-Wave Theory}
\label{app:lswt_results}

The main text focuses on DSSF data generated using THL. For comparison, this appendix collects the corresponding results obtained using LSWT. LSWT provides a useful controlled benchmark because the forward map is computationally inexpensive and the resulting spectra contain predominantly single-magnon excitations. The LSWT spectra in this appendix are computed at the field $B_x=B_\mathrm{sat}(\mathbf{J})+|J_1|$, as in the main text.

\subsection{Tensor components in LSWT}

The corresponding LSWT tensor components differ from the THL results in Fig.~\ref{fig:distinguishable}. In LSWT, only $S_{yz}+S_{zy}$ reverses sign with $J_{z\pm}$, while $S_{xy}+S_{yx}$ and $S_{zx}+S_{xz}$ vanish. Along $(h,h,1)$, LSWT therefore still distinguishes the two signs. Along the accessible cut $(2h,-h,1)$, however, $\widehat q_y=0$ removes $S_{yz}+S_{zy}$ from the unpolarized intensity, and the sign can be resolved only when the forward model includes the mixing of one- and two-magnon states.

\subsection{LSWT identifiability}
\label{app:lswt_identifiability}

We first examine the identifiability of the Hamiltonian parameters using the FIM.
The four-parameter model is
\[
\mathbf{J} =
(J_1,J_{zz},J_{\pm\pm},J_{z\pm},J_2)
=(1,0.2,0.1,0,0.5),
\]
whereas the five-parameter model is
\[
\mathbf{J} =
(1,0.2,0.1,0.05,0.5).
\]
The LSWT results are summarized in Table~\ref{apptab:app_lswt_identifiability}.
The addition of the full-tensor DSSF substantially improves the conditioning of the five-parameter inversion, particularly when the off-diagonal coupling $J_{z\pm}$ is included.

\begin{table}[t]
\caption{Identifiability of four- and five-parameter Hamiltonians using LSWT.
The mean squared error (MSE) and Fisher-information (FI) condition number $\kappa$ are reported for unpolarized and polarized DSSF as well as different momentum cuts $(HHQ_z)$, where $Q_z$ is the out-of-plane momentum component. $Q_z=0$ corresponds to the in-plane cut. The four-parameter model excludes $J_{z\pm}$, whereas the five-parameter model includes it. The data use $N_k=20$ momentum points and $N_J=20$ Hamiltonian samples. The inverse model is the Conv1D network with hidden dimension $64$ and trained with 100 epochs. ``Full tensor'' is defined as in Table~\ref{tab:identifiable}, and the $(HH1)$ and $(HH2)$ cuts are an idealized geometry.}
\label{apptab:app_lswt_identifiability}
\centering
\begin{ruledtabular}
\begin{tabular}{lcc}
Model / DSSF & MSE & $\kappa$ \\
\hline
\multicolumn{3}{c}{$4$ parameters ($J_{z\pm}$ excluded)} \\
\hline
Unpolarized, $(HH0)$
    & $1.45\times10^{-5}$ & $5.22\times10^{1}$ \\
Full Tensor, $(HH0)$
    & $9.98\times10^{-6}$ & $5.10\times10^{1}$ \\
Unpolarized, $(HH1)$
    & $1.28\times 10^{-5}$ & $5.23\times10^{1}$ \\
Unpolarized, $(HH2)$
    & $1.37\times10^{-5}$ & $5.24\times10^{1}$ \\
\hline
\multicolumn{3}{c}{$5$ parameters ($J_{z\pm}$ included)} \\
\hline
Unpolarized, $(HH0)$
    & $7.73\times10^{-2}$ & $1.68\times10^{5}$ \\
Full Tensor, $(HH0)$
    & $1.43\times10^{-5}$ & $1.73\times10^{3}$ \\
Unpolarized, $(HH1)$
    & $1.22\times 10^{-5}$ & $1.24\times10^{4}$ \\
Unpolarized, $(HH2)$
    & $1.43\times 10^{-5}$ & $4.92\times10^{3}$ \\
\end{tabular}
\end{ruledtabular}
\end{table}

\subsection{Conditioning near vanishing off-diagonal exchange}
\label{app:conditioning_jzpm}

To examine how the off-diagonal exchange affects local Hamiltonian inference, we evaluate the Fisher-information condition number $\kappa=\lambda_{\max}/\lambda_{\min}$ as a function of $J_{z\pm}$, holding the remaining Hamiltonian parameters fixed. We compare the THL calculation with LSWT using the in-plane $Q_z=0$ unpolarized DSSF. The results are summarized in Table~\ref{apptab:conditioning_jzpm}.

\begin{table}[t]
\caption{Fisher-information condition number $\kappa$ versus $J_{z\pm}$ for the THL and LSWT calculations with $\mathbf{J}=(J_1,J_{zz},J_{\pm\pm},J_{z\pm},J_2)
=(1,0.7,0.2,J_{z\pm},0.051)$. The results for positive and negative couplings agree to the displayed precision and are therefore grouped by $|J_{z\pm}|$. At zero coupling, the ladder calculation returns an infinite condition number, whereas LSWT returns a very large finite value, consistent with numerical singularity. The number of momentum points $N_k=60$ and finite difference $10^{-3}$.}
\label{apptab:conditioning_jzpm}
\centering
\begin{ruledtabular}
\begin{tabular}{ccc}
$|J_{z\pm}|$ & $\kappa$, THL & $\kappa$, LSWT \\
\hline
$0$    & $\infty$             & $1.49\times10^{25}$ \\
$0.05$ & $5.62\times10^{4}$  & $1.84\times10^{5}$ \\
$0.10$ & $1.43\times10^{4}$  & $4.60\times10^{4}$ \\
$0.20$ & $3.81\times10^{3}$  & $1.16\times10^{4}$ \\
\end{tabular}
\end{ruledtabular}
\end{table}

Both calculations become increasingly ill-conditioned as $J_{z\pm}$ approaches zero. The THL calculation has a smaller condition number than LSWT at each nonzero coupling, by a factor of approximately three under the parameter scaling and observation model used here, indicating an improved local identifiability due to the many-body effect.

This behavior is consistent with the sign ambiguity of the in-plane unpolarized response discussed in the main text. Improved conditioning at nonzero $|J_{z\pm}|$ does not establish global identifiability. A response that is even in $J_{z\pm}$ can locally constrain its magnitude while remaining unable to distinguish its sign. Resolving that ambiguity requires an observable sensitive to the sign, such as the cross-correlation channel or an appropriate out-of-plane geometry discussed in the main text.

\subsection{Neural-network architecture benchmark}
\label{app:lswt_architectures}

Table~\ref{apptab:lswt_architectures} compares several neural-network architectures on the LSWT
inversion task.
The models use the same input representation and hidden dimension; only the architecture is varied.
The performance differences are small, indicating that the LSWT inverse map is relatively smooth and
can be learned effectively by a compact regressor.

\begin{table}[t!]
\caption{Neural-network performance on LSWT-generated unpolarized DSSF (above two lines: clean data,
below two lines: noisy data with $\sigma_\epsilon=0.1$) for a four-parameter Hamiltonian and
$N_J=30, N_k=250$.
MSE and MARE are from testing data results.
Neural networks with hidden dimension $64$ are used.}
\label{apptab:lswt_architectures}
\centering
\small
\resizebox{\columnwidth}{!}{%
\begin{ruledtabular}
\begin{tabular}{lcccccc}
Metric & FCNN & Conv1D & Transformer & ConvTransformer \\
\hline
MSE & $3.29\times10^{-5}$ & $\mathbf{1.43\times10^{-5}}$ & $2.28\times10^{-4}$ & $3.75\times10^{-5}$ \\
MARE & $1.04\times10^{-2}$ & $\mathbf{8.61\times10^{-3}}$ & $2.96\times10^{-2}$ & $1.43\times10^{-2}$ \\
\hline
MSE & $1.44\times10^{-3}$ & $\mathbf{1.23\times10^{-4}}$ & $1.40\times10^{-3}$ & $1.38\times10^{-3}$ \\
MARE & $8.15\times10^{-2}$ & $\mathbf{7.33\times10^{-2}}$ & $7.60\times10^{-2}$ & $7.79\times10^{-2}$ \\
\end{tabular}
\end{ruledtabular}
}
\end{table}

Fig. \ref{appfig:lswt_error_plot} shows the corresponding parity plot of the predicted Hamiltonian parameters versus the true Hamiltonian parameters, as a counterpart to Fig. \ref{fig:error_plot} in the main text. In the LSWT case, the inversion is simpler than for THL data, and the neural network achieves a tiny MARE without noise augmentation. 

\begin{figure}[t!]
\begin{center}
\includegraphics[width=\columnwidth]{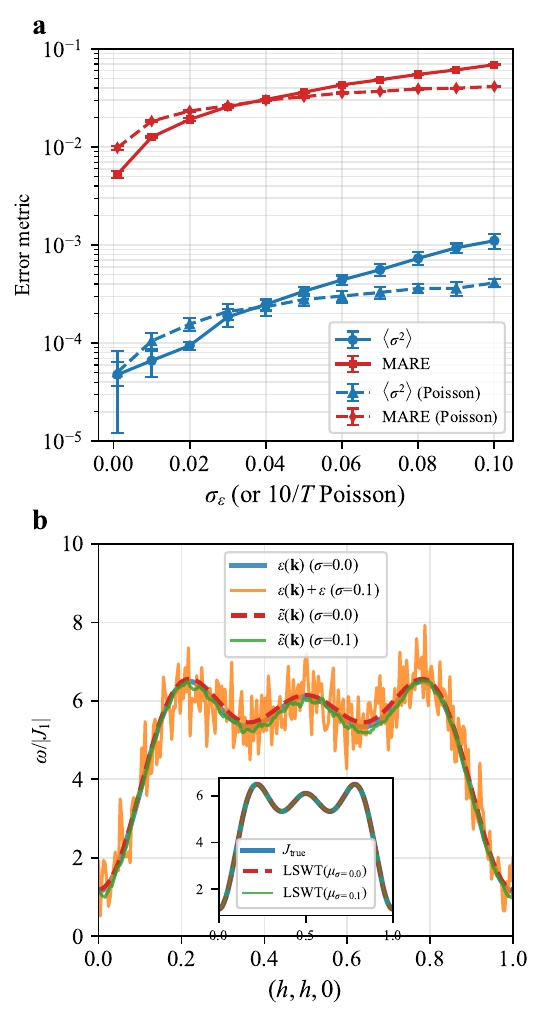}
\caption{\label{appfig:noise} 
The effect of augmenting the training data with noise on the performance of the neural network for LSWT data ($N_k=250,N_J=20$). Top panel: the mean predicted variance by the neural network and the mean absolute relative error (MARE) of the predicted Hamiltonian parameters as a function of the noise level $\sigma_\epsilon$ (or Poisson noise $10/T$, which is proportional to the Poisson variance) added to the DSSF data during training. Error bars are standard deviation from 20 training epochs. Bottom panel: comparison of the predicted DSSF data (dispersion only $\tilde{\varepsilon}(\mathbf{k})$) from the neural network with the true DSSF dispersion ($\varepsilon(\mathbf{k})$) for a test sample $\mathbf{J}=(-1,-0.9,-0.8,0,-0.5)$. The predicted Hamiltonian parameters are $\mu_{\sigma=0.0}(\mathbf{J})=(-1.01(1), -0.909(5), -0.800(9), 0, -0.496(5))$ and $\mu_{\sigma=0.1}(\mathbf{J})=(-0.99(1), -0.88(4), -0.80(5), 0, -0.51(2))$, where the error is given by one standard deviation. Inset: the true DSSF dispersion versus the predicted DSSF dispersion from LSWT using the predicted mean Hamiltonian parameters $\mu_{\sigma}(\mathbf{J})$. 
}
\end{center}
\end{figure}

Fig. \ref{appfig:noise} shows the effect of augmenting the training data with noise on the
performance of the neural network for LSWT data.
The Poisson curves are parameterized by $1/T$, which controls intensity variance,
whereas $\sigma_\epsilon$ controls the standard deviation of the Gaussian perturbations.
The bottom panel shows the performance of the decoder for a test sample.
Even with strong noise, the reconstructed dispersion remains close to the true dispersion. 

\subsection{Convergence with respect to the number of uniform parameter
points}\label{app:lswt_convergence}

We examine the convergence of the Conv1D as the number of uniformly sampled parameter points $N_J$ is increased. The results are summarized in Table~\ref{apptab:lswt_convergence}. The test MSE and MARE show non-monotonic behavior with increasing $N_J$. The saturated performance is obtained for $N_J=20$, suggesting that increasing the number of training samples beyond this value does not necessarily improve the inversion accuracy under the present training settings.

For the THL solver, limited by the memory (data with many energy points per momentum), we focus on the saturated number of training samples, $N_J=20$ in the main text. The test MARE converges to around $2\%$, which is already a good performance for the inversion task.  

\begin{table}[t!]
\caption{Convergence of the Conv1D with respect to the number of uniformly sampled parameter points $N_J$ with $N_k=20$. The test MSE and MARE are reported for the LSWT-generated dataset with four Hamiltonian parameters.}
\label{apptab:lswt_convergence}
\centering
\resizebox{\columnwidth}{!}{%
\begin{ruledtabular}
\begin{tabular}{lccccc}
Metric & $N_J=10$ & $N_J=20$ & $N_J=30$ & $N_J=50$ \\
\hline
MSE
& $1.64\times10^{-5}$
& $\mathbf{1.45\times10^{-5}}$
& $1.63\times10^{-5}$
& $9.83\times10^{-6}$ \\
MARE
& $5.97\times10^{-3}$
& $\mathbf{7.01\times10^{-3}}$
& $8.75\times10^{-3}$
& $7.06\times10^{-3}$ \\
\end{tabular}
\end{ruledtabular}
}
\end{table}

\subsection{Dependence on number of momenta and energy bins}
\label{app:lswt_resolution}

We next study how the number of retained momentum and energy points affects the inversion accuracy.
Table~\ref{apptab:nk_length_noise} shows the dependence on the number of momentum points for both
clean and noisy LSWT data.
The results indicate that the neural network can learn the inverse map effectively even with a
relatively small number of momentum points, i.e., $N_k=20$ in the clean case.
A larger value of $N_k$ actually leads to slightly worse performance,
which is due to the increased input size and the limited model capacity (fixed hidden dimension and
training epochs).
On the other hand, the noisy data case shows a clear improvement with increasing $N_k$,
as the additional momentum points help to average out the noise and provide more information for the
regression task.
With noise of fixed amplitude per momentum point, the error decreases steadily as $N_k$ grows (Table~\ref{apptab:nk_length_noise}). In a counting experiment, however, more momentum points at the same noise level per point require proportionally more counts; at fixed exposure, finer binning only raises the noise in each bin (Sec.~\ref{sec:resolution} and Appendix~\ref{app:fisherinfo}). The improvement therefore reflects additional total counts rather than finer sampling of a fixed measurement, and models intended for noisy data should be trained with a noise model matched to the expected counting statistics, as discussed in \cite{luo2026approaching}.

\begin{table*}[t]
\caption{Neural network (Conv1D with hidden dimension $64$) performance on unpolarized DSSF with 4 Hamiltonian parameters versus number of $k$-points ($N_k$) trained without noise $\sigma=0$ (top two lines) and with noise $\sigma=0.1$ (bottom two lines). The density of data is $N_J=20$.}
\label{apptab:nk_length_noise}
\begin{ruledtabular}
\begin{tabular}{lcccccc}
Metric & 20 & 50 & 100 & 250 & 500 & 1000 \\
\hline
MSE & $1.454\times10^{-5}$ & $1.633\times10^{-5}$ & $2.983\times10^{-5}$ & $2.715\times10^{-5}$ &
$3.895\times10^{-5}$ & $3.176\times10^{-5}$  \\
MARE  & $7.013\times10^{-3}$ & $7.256\times10^{-3}$ & $1.117\times10^{-3}$ & $8.277\times10^{-3}$ &
$1.246\times10^{-2}$ & $1.014 \times10^{-3}$ \\
\hline
MSE  & $1.104\times10^{-2}$ & $5.747\times10^{-3}$ & $3.094\times10^{-3}$ & $1.546\times10^{-3}$ &
$1.306\times10^{-3}$ & $1.002\times10^{-3}$\\
MARE  & $1.924\times10^{-1}$ & $1.384\times10^{-1}$ & $1.037\times10^{-1}$ & $7.185\times10^{-2}$ &
$6.354\times10^{-2}$ & $5.713\times10^{-2}$ \\
\end{tabular}
\end{ruledtabular}
\end{table*}

\subsection{Classical regression baselines}
\label{app:lswt_classical_baselines}

For LSWT-generated data, classical nonlinear regression methods can perform particularly well
because the forward map is smooth and inexpensive to evaluate.
The comparison in Table~\ref{apptab:lswt_classical_baselines} shows that Gaussian-process regression
(GPR) gives the lowest test error among the methods considered.
Its computational cost, however, increases rapidly with the size of the training set,
making neural networks more attractive for larger datasets or repeated inference.

\begin{table*}[t]
\caption{Classical regression baseline performance on unpolarized DSSF inversion (4 Hamiltonian parameters, $N_k=20, N_J=20$) under clean scalar data using LSWT.}
\label{apptab:lswt_classical_baselines}
\centering
\footnotesize
\begin{ruledtabular}
\resizebox{\textwidth}{!}{%
\begin{tabular}{lccccccccc}
Metric & Ridge & ElasticNet & PLS & Kernel Ridge & SVR & \textbf{GPR (cap 2000)} & Random Forest &
Gradient Boosting & k-Nearest Neighbor \\
\hline
MSE     & $1.79\times10^{-2}$ & $2.67\times10^{-2}$ & $1.46\times10^{-2}$ & $2.28\times10^{-4}$ & $2.74\times10^{-5}$ & $\mathbf{4.82\times10^{-11}}$ & $8.84\times10^{-4}$ & $7.74\times10^{-4}$ & $5.19\times10^{-4}$ \\
MARE  & $1.83\times10^{-1}$ & $2.11\times10^{-1}$ & $1.66\times10^{-1}$ & $1.76\times10^{-2}$ & $1.06\times10^{-2}$ & $\mathbf{5.54\times10^{-6}}$ & $5.12\times10^{-2}$ & $4.90\times10^{-2}$ & $4.84\times10^{-2}$ \\
\end{tabular}
}
\end{ruledtabular}
\end{table*}

The fact that for LSWT-generated data, the GPR method achieves the lowest test MARE (3 orders of magnitude lower than the next best method) indicates that the LSWT forward map under single-magnon excitations can be effectively learned by a smooth kernel-based regression. Physically, the relation between the Hamiltonian parameters and the LSWT band may have simple functional forms at high-symmetry momentum points, which can be captured by the GPR method. We conjecture that the GPR method may perform exceptionally well for the inversion of generic band-like spectra, such as band structures from single-particle tight-binding models.

When adding noise to the LSWT data, the GPR method is no longer the best-performing method. But its MSE and MARE remain comparable to those of the neural network methods. The drawback of the GPR method is that it is not scalable to large datasets, as the computational cost scales cubically with the number of training samples \cite{williams2006gaussian}. In contrast, neural networks can be trained on larger datasets and can generalize well to unseen data. 

The GPR works well for LSWT-generated data because the forward map is smooth and the spectra are relatively simple. For THL-generated data, the forward map is more complex and the spectra contain multi-particle excitations, which makes the inversion task more challenging. In this case, neural networks are more suitable due to their ability to learn complex mappings from high-dimensional inputs to outputs.

\clearpage
\bibliography{refs.bib}

\end{document}